\documentclass[acmsmall,screen,nonacm]{acmart}

\usepackage{subcaption}
\usepackage{caption}
\usepackage{graphicx}
\usepackage{multirow}
\usepackage{appendix}
\usepackage{tikz}
\usepackage[dvipsnames]{xcolor}
\usepackage{pgfplots}
\usepgfplotslibrary{groupplots}
\usetikzlibrary{patterns,patterns.meta,positioning}
\pgfplotsset{compat=1.18}
\usepackage{model_icons}
\AtBeginDocument{
  }

\begin{document}

\title{AI Code in the Wild: Measuring Security Risks and Ecosystem Shifts of AI-Generated Code in Modern Software}
\author{Bin Wang}
\affiliation{%
  \institution{Peking University}%
  \country{China}%
}
\email{thebinking66@stu.pku.edu.cn}

\author{Wenjie Yu}
\affiliation{%
  \institution{Peking University}%
  \country{China}%
}
\email{uuykea@gmail.com}

\author{Yilu Zhong}
\affiliation{%
  \institution{Peking University}%
  \country{China}%
}
\email{tangaaang@gmail.com}

\author{Hao Yu}
\affiliation{%
  \institution{Peking University}%
  \country{China}%
}
\email{g.diorld@gmail.com}

\author{Keke Lian}
\affiliation{%
  \institution{Tencent}%
  \country{China}%
}
\email{kekelian@tencent.com}

\author{Chaohua Lu}
\affiliation{%
  \institution{Tencent}%
  \country{China}%
}
\email{ciuwaalu@tencent.com}

\author{Hongfang Zheng}
\affiliation{%
  \institution{Tencent}%
  \country{China}%
}
\email{koalazheng@tencent.com}

\author{Dong Zhang}
\authornote{Corresponding author.}
\affiliation{%
  \institution{Tencent}%
  \country{China}%
}
\email{zalezhang@tencent.com}

\author{Hui Li}
\authornotemark[1]
\affiliation{%
  \institution{Peking University}%
  \country{China}%
}
\email{lih64@pkusz.edu.cn}

\renewcommand{\shortauthors}{Wang et al.}

\begin{abstract}
Large language models (LLMs) for code generation are becoming integral to modern software development, but their real-world prevalence and security impact remain poorly understood.

We present the first large-scale empirical study of AI-generated code (AIGCode) in the wild. We build a high-precision detection pipeline and a representative benchmark to distinguish AIGCode from human-written code, and apply them to (i) development commits from the top 1,000 GitHub repositories (2022-2025) and (ii) 7,000+ recent CVE-linked code changes. This lets us label commits, files, and functions along a human/AI axis and trace how AIGCode moves through projects and vulnerability life cycles.

Our measurements show three ecological patterns. First, AIGCode is already a substantial fraction of new code, but adoption is structured: AI concentrates in glue code, tests, refactoring, documentation, and other boilerplate, while core logic and security-critical configurations remain mostly human-written. Second, adoption has security consequences: some CWE families are overrepresented in AI-tagged code, and near-identical insecure templates recur across unrelated projects, suggesting "AI-induced vulnerabilities" propagated by shared models rather than shared maintainers. Third, in human-AI edit chains, AI introduces high-throughput changes while humans act as security gatekeepers; when review is shallow, AI-introduced defects persist longer, remain exposed on network-accessible surfaces, and spread to more files and repositories.

We will open-source the complete dataset and release analysis artifacts and fine-grained documentation of our methodology and findings.

\end{abstract}


\maketitle

\section{Introduction}

With the rapid development of large-scale language models, artificial intelligence technology is profoundly reshaping the field of software engineering. Code generation tools, such as GitHub Copilot\cite{yeticstiren2023evaluating,pandey2024transforming} and Claude Code\cite{nettur2025role,dong2025survey}, have demonstrated stable and significant performance improvements in areas such as code completion, function implementation, refactoring, and project-level task collaboration \cite{chen2021evaluating,lyu2025automatic,peng2023impact,song2024impact,asraful2024llms,wang2025argus}. Statistics show that a considerable proportion of developers are already using AI tools to write code in their daily work \cite{hajipour2024codelmsec, dilgren2025secrepobench}, and AIGCode has become a crucial component of the modern software development process.

However, this technological shift also brings potential security risks. LLMs' power stems from their learning from massive amounts of publicly available codebases, but this data inevitably includes historical code with security flaws\cite{hajipour2024codelmsec, siddiq2022securityeval,lian2025aserepositorylevelbenchmarkevaluating,wang2025promptpoisoningcodedefect}. Therefore, the model inherits these insecure coding patterns during the learning process, potentially causing the generated code to reproduce known vulnerabilities\cite{siddiq2022securityeval, he2024instruction} or even introduce new, more subtle security risks when interacting with context\cite{dilgren2025secrepobench, ahmed2025secvuleval}. Furthermore, the convenience and efficiency of AIGCode allow for the generation and revision of large amounts of code at once through developer-guided conversations, often without careful review and security checks, further increasing the probability of code vulnerabilities.

While academia and industry have directed attention to the security issues of AIGCode\cite{fu2025security,Wang_2025,liu2025ra}, a fundamental deficiency persists in existing research: the lack of systematic and large-scale empirical studies on AIGCode in the wild. This knowledge gap prevents a comprehensive understanding of AIGCode's actual penetration scale, specific security impact, and long-term risk evolution trends in real-world software projects. This insufficient empirical basis, in turn, severely restricts the effective formulation and deployment of risk mitigation strategies.The primary technical obstacle to conducting such empirical analysis is the challenge of code provenance and detection. Since AIGCode is highly similar to human-written code in syntax, style, and logic, there is a lack of effective technical means to accurately distinguish it, making effective regulation, risk assessment, and liability determination extremely difficult.

To bridge this methodological gap and enable systematic empirical analysis,  this paper proposes an ensemble learning framework for AIGCode detection and performs a comprehensive empirical analysis of GitHub projects to dissect the associated trends and security risks. We designed and implemented an AIGCode detection framework, which innovatively adopts an ensemble learning strategy to improve detection accuracy and speed by combining the strengths of multiple base models. Based on this, we conducted a large-scale empirical study on publicly available vulnerability intelligence and the codebase of GitHub's Top 1,000 open-source projects, aiming to systematically reveal the current penetration status of AIGCode in the wild and analyze its potential correlation with security vulnerabilities.

The main contributions of this paper are summarized as follows:
\begin{enumerate}
\item \textbf{A Novel Detection Framework.} We designed and implemented an efficient, ensemble learning-based AIGCode detection framework, Cascade-Aggregation Framework. Experimental results show that this framework demonstrates state-of-the-art accuracy and robustness in distinguishing between real-world AI and human code. We will open-source the corresponding detection model and dataset.

\item  \textbf{The First Large-Scale Ecosystem Analysis.} We conducted the first large-scale empirical study on the Top 1000 open-source projects ranked by GitHub stars, quantitatively analyzing the scale and penetration trend of AIGCode in real-world software development.

\item \textbf{A Comprehensive Security Risk Profile.} By performing correlation analysis between vulnerability intelligence and AIGCode usage in recent years, we construct a detailed security risk profile for AIGCode, revealing its unique risk patterns.
\end{enumerate}

\section{RELATED WORKS}

\subsection{AIGCode Code Detection}

With the rapid popularity of LLM-driven programming assistance tools, distinguishing AIGCode from human-written code has garnered growing attention in software security and trusted computing\cite{orel2025droid,xu2024detecting,bulla2024ex,patel2024comparative,demirok2025multiaigcd}. Existing methods are broadly divided into two categories: \textbf{proactive provenance} and \textbf{passive content analysis}\cite{guan2024codeip,guo2025codemirage}.

\paragraph{Proactive Provenance} This category, typified by digital watermarks and signatures, embeds verifiable ``fingerprints'' directly within the code during the generation phase, thereby theoretically offering high verifiability and robustness\cite{guan2024codeip,zhang2025robust,zhao2411sok}. However, this approach relies on the cooperation and modification of the model or server, making it difficult to cover the large amount of existing \emph{in-the-wild} code, and also difficult to trace historical data and fragments after cross-platform copying. Therefore, it is not suitable for our task scenario aimed at identifying AIGCode in the existing ecosystem\cite{guo2025codemirage,suh2024empirical}.

\paragraph{Passive Content Analysis}
Starting from the generated content itself, this approach can typically be divided into three categories: (1) \emph{Statistical and Style Features}: Using statistical measures such as complexity\cite{imperial2023uniform,doughman2025exploring}, token spectrum or metrics like entropy\cite{wu2025survey,civico2025linguistic} or perplexity\cite{megias2024influence}, as features for discrimination, typically combined with traditional classifiers\cite{nguyen2024gptsniffer}; (2) \emph{Representation Learning Discrimination}: Utilizing pre-trained code representation models to represent code snippets and perform binary classification\cite{feng2020codebert,ma2024enhancing}. \textsc{EX-CODE} achieves detection by estimating the occurrence probability of different types of tokens and combining them with a classifier, and also possesses a certain degree of interpretability\cite{bulla2024ex}; (3) \emph{Structural and Semantic Constraints}: Integrating structural features such as Abstract Syntax Trees , Control Flow Graphs, or Intermediate Representation with semantic signals such as API usage and code smells. Alternatively, it introduces lightweight execution and unit testing behavioral features to enhance the cross-language and cross-style robustness of the discrimination\cite{pan2024assessing,cotroneo2024vulnerabilities,tihanyi2025hidden,yang2022natural,guan2024codeip}.
Although this approach has made progress, it still faces common limitations such as \emph{insufficient generalization}, \emph{adversarial vulnerability}, and \emph{contextual myopia}\cite{cotroneo2024vulnerabilities,shukla2025security,sabra2025assessing,tihanyi2025hidden}. 
\subsection{Security of AIGCode}

The widespread proliferation of tools like GitHub Copilot and CodeX \cite{chen2021evaluating,hajipour2024codelmsec,dilgren2025secrepobench} has brought the security issues of AIGCode into sharp focus\cite{ji2024cybersecurity,wang2025mcpguardautomaticallydetecting,taeb2024assessing,cotroneo2024vulnerabilities,fu2025security,hajipour2024codelmsec,dilgren2025secrepobench}. Models may replicate insecure patterns from training data or introduce insecure dependencies\cite{ji2024cybersecurity,cotroneo2024vulnerabilities,sabra2025assessing,fu2025security,bashir2025using,siddiq2022securityeval,hajipour2024codelmsec}, leading to \emph{exploitable vulnerabilities} and \emph{supply chain risks}\cite{shukla2025security}.

\paragraph{Risk Characterization and Empirical Studies} Several studies report that a significant proportion of AIGCode contains security flaws\cite{cotroneo2024vulnerabilities,sabra2025assessing,patel2024comparative,tihanyi2025hidden,hamer2024just,perry2023users}, frequently occurring in weak areas such as input validation and access control \cite{ji2024cybersecurity,taeb2024assessing,sandoval2022security,perry2023users}. Recent work attempts to construct an ``AIGCode risk profile'' from dependency security, call graphs, and behavioral characteristics to quantify security features \cite{shukla2025security,fu2025security,ahmed2025secvuleval}.

\paragraph{Detection and Protection}Given the insufficiency of traditional static and dynamic analysis in covering semantic and logical defects, researchers have introduced LLMs to enhance semantic-level detection capabilities\cite{yang2022natural,temtsin2025imitation,suh2024empirical,gao2023far}. On the defense side, one approach applies security alignment and prompt constraints \emph{before generation} to reduce risks \cite{shukla2025security,zhao2411sok,homoliak2024enhancing}; another performs automatic auditing and repair \emph{after generation}, combining static scanning with automatic patching\cite{fu2025security,bashir2025using,he2024instruction,kim2024codexity}.
However, current research is still limited by \emph{a lack of evaluation benchmarks and real-world ecological data}\cite{guo2025codemirage,suh2024empirical}, \emph{insufficient generalization across languages and novel vulnerabilities}\cite{cotroneo2024vulnerabilities,tihanyi2025hidden}, and \emph{a lack of end-to-end verifiable metrics}\cite{taeb2024assessing,guo2025codemirage,oedingen2024chatgpt,dilgren2025secrepobench,ahmed2025secvuleval}.

\section{AIGCodeDetector}

Current research on AIGCode security generally lacks accurate identification and quantifiable measurement of code sources in real-world development ecosystems. Existing work often relies on synthetic data in controlled environments and single-dimensional statistical features, resulting in insufficient external validity and difficulty in robustly transferring conclusions to complex open-source code repositories in the wild. To fill this methodological gap and support our systematic research on the prevalence of AIGCode in in-the-wild repositories and its correlation with security vulnerabilities, This section introduces the Cascade–Aggregation Framework, which we design and implement in this work.

\subsection{Construction of the AIGCode Detection Evaluation Dataset}
To systematically evaluate the robustness and cross-domain generalization ability of AIGCode detection models in real-world scenarios, we constructed a large-scale, high-confidence evaluation benchmark dataset. This dataset consists of two complementary subsets: the \emph{trusted human code subset} and the \emph{trusted AIGCode subset}. The overall design follows three core principles: \textbf{label purity} (avoiding annotation confusion and errors), \textbf{sample authenticity} (closely reflecting real-world engineering environments), and \textbf{scenario diversity} (covering multiple programming languages, task types, and application domains).

\paragraph{Trusted Human Code Subset}
This subset aims to provide \emph{high-quality human-written code samples unaffected by AI programming tools}. To ensure the purity of the labels, we limited the data source to code commits on GitHub between \textbf{January 1, 2008} and \textbf{December 31, 2010}. This time window predates the rise of large language model-based code generation tools, and traditional AI-assisted programming methods were not widely used either, thus fundamentally guaranteeing the reliability of the \texttt{Human-written} label.

In the specific construction process, we crawled code from a set of highly-starred core open-source repositories for each of the 10 commonly used programming languages within the selected time period, ensuring that the collected samples exhibit industrial-grade quality and strong representativeness. Subsequently, we filtered out non-code files such as documentation and configuration files using file extensions, retaining only source code files containing executable logic. For efficient model ingestion and batch processing, the code's provenance, the code content, and the corresponding ground-truth label are serialized. This pipeline converts unstructured historical version data into standardized, file-granularity evaluation instances.

The human code corpus contains approximately 40,000 code files, covering 10 mainstream programming languages. Among them, Java, C++, and Python account for the highest proportions, mainly corresponding to typical scenarios such as enterprise-level development, system components, and scientific computing. JavaScript and C\# are widely used in front-end and back-end applications as well as engineered software development. Ruby and PHP still exist in traditional Web businesses. While languages like Go, Rust, and Shell are smaller in scale, they have representative value in distributed systems and engineering operations.

\paragraph{Trusted AIGCode Subset.} This subset is designed to provide highly challenging and structured AIGcode for evaluation tasks. We employed a structured synthesis methodology to construct this subset, aiming to faithfully simulate AI code generation behaviors, particularly those encountered in safety-critical development environments. This process rigorously tests the detection model's capability to accurately identify the distinct characteristics of machine-generated code.

We first designed 33 common programming topics, further refining them into 165 specific programming tasks, ensuring that the generated code samples possess sufficient diversity and representativeness in terms of functional complexity, logical structure, and application scenarios. To verify the cross-model generalization ability of the detection model, we used 11 mainstream large language models for code generation, including closed-source models and open-source models, thus covering the differences in generation styles brought about by different LLM architectures and training strategies.

\paragraph{Dataset Profile.}
The evaluation dataset encompasses ten mainstream programming languages across nine major domains, including system software, web applications, and data science. Such broad language and domain coverage ensures a robust basis for assessing model discriminative capability under diverse code styles and complexities.

\begin{figure*}[h]
\centering
\begin{subfigure}[t]{0.24\textwidth}
    \centering
    \includegraphics[width=1.0\linewidth]{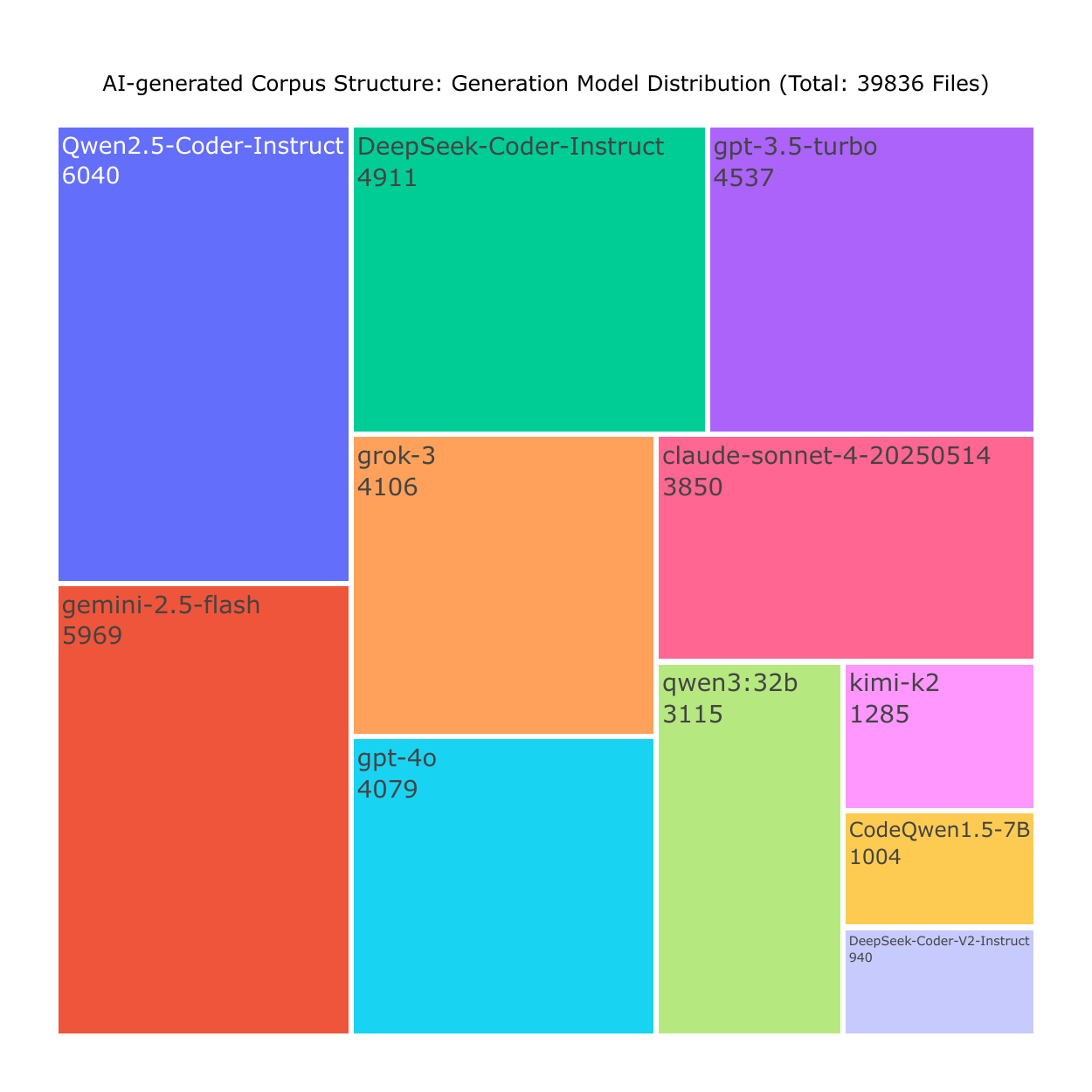} %
    \subcaption{}
    \label{fig:ai_model_treemap}
\end{subfigure}
\hfill
\begin{subfigure}[t]{0.24\textwidth}
    \centering
    \includegraphics[width=1.0\linewidth]{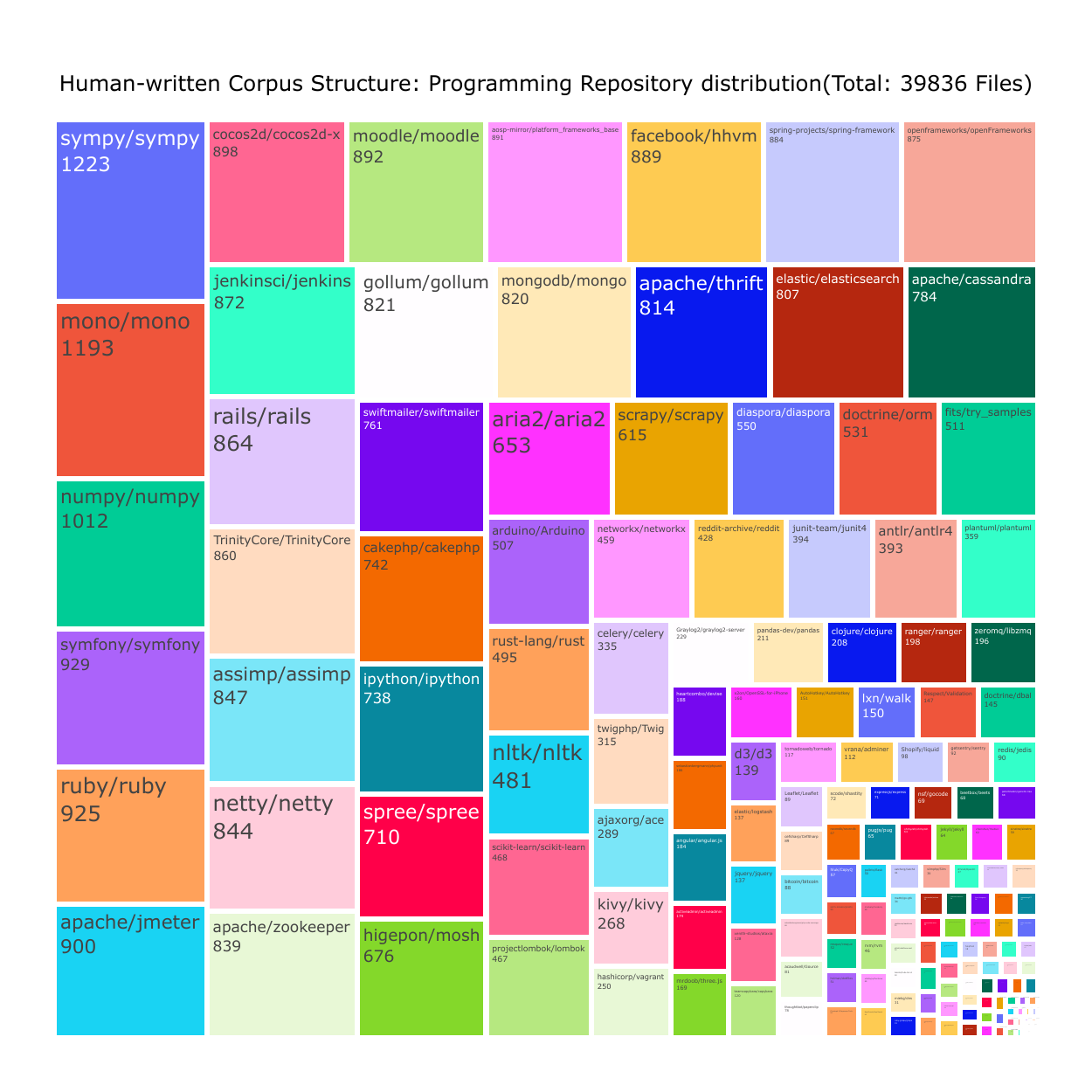}
    \subcaption{}
    \label{fig:human_repo_treemap}
\end{subfigure}
\hfill
\begin{subfigure}[t]{0.24\textwidth}
    \centering
    \includegraphics[width=1.0\linewidth]{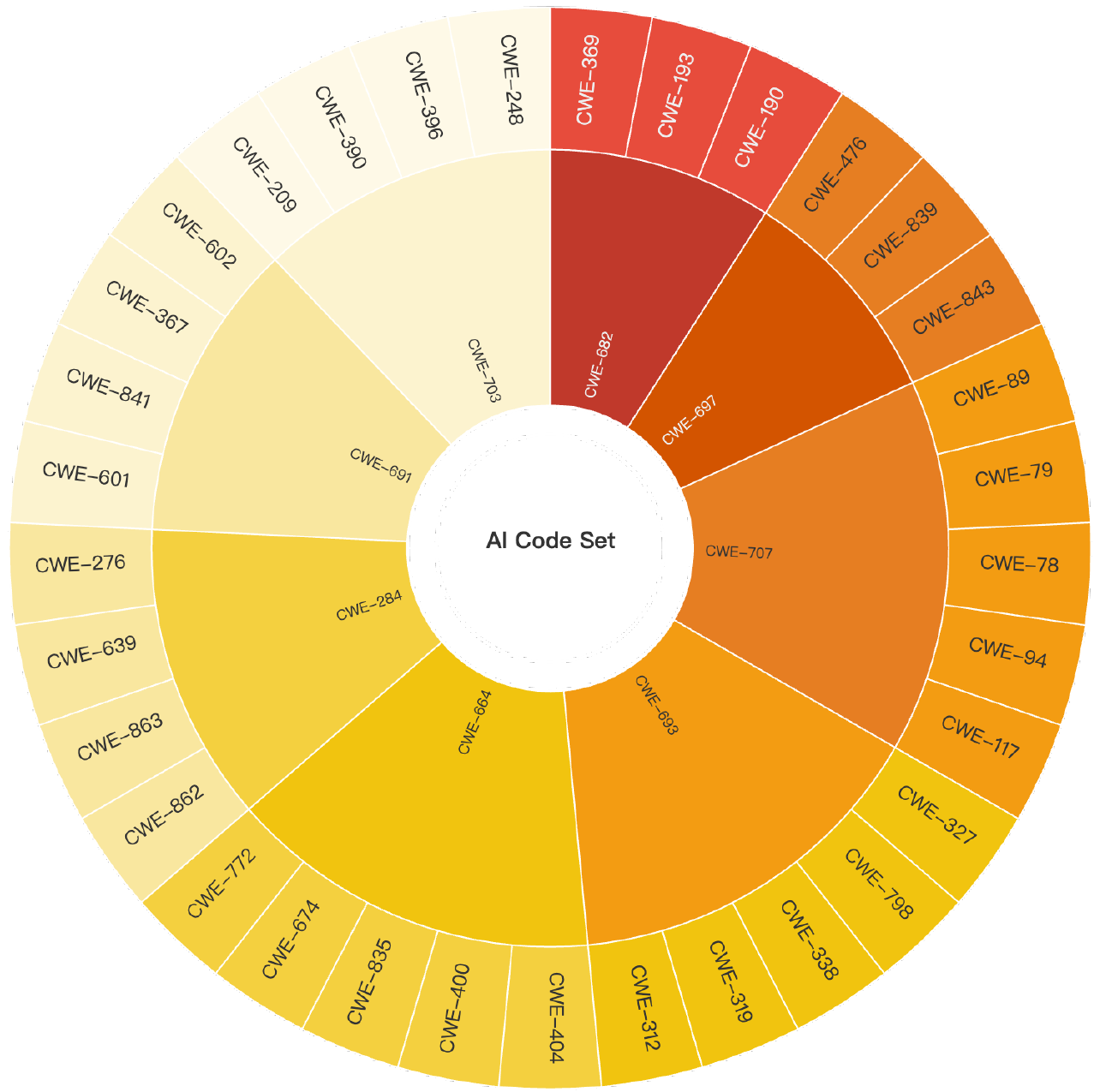}
    \subcaption{}
    \label{fig:Dataset_CWE_Distribution}
\end{subfigure}
\hfill
\begin{subfigure}[t]{0.24\textwidth}
    \centering
    \includegraphics[width=1.0\linewidth]{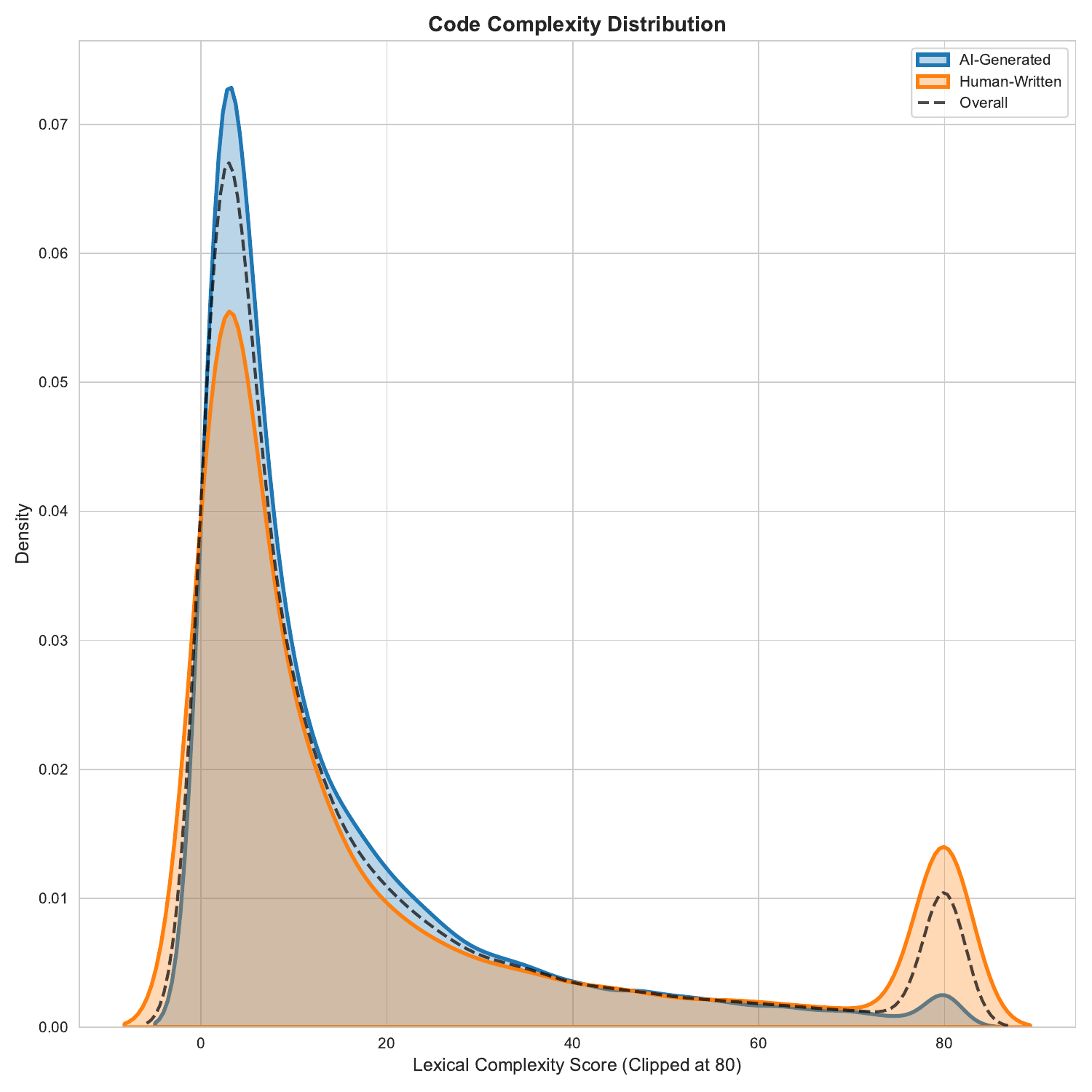}
    \subcaption{}
    \label{fig:Code Complexity Distribution}
\end{subfigure}

\caption{Dataset Characterization. 
(a) Distribution of the AI-generated corpus by generation model. 
(b) Distribution of the Human-written corpus by programming repository. 
(c) Dataset CWE Distribution. 
(d) Code Complexity Distribution.}
\label{fig:dataset_distributions}

\end{figure*}

The dataset consists of two balanced subsets of 40K samples each. The human-written subset is collected from 172 open-source repositories, primarily featuring Java, C++, and Python projects from enterprise, system, and automation scenarios. As illustrated in Fig.\ref{fig:dataset_distributions}(\subref{fig:human_repo_treemap}), it exhibits a long-tailed distribution—large, mature projects provide well-structured modules, while smaller repositories contribute diverse coding styles. The AI-generated subset is synthesized using both proprietary and open-source models, as detailed in Fig.\ref{fig:dataset_distributions}(\subref{fig:ai_model_treemap}).

We categorized the samples by application domain (Tab.\ref{tab:Application Domain}). Code related to Web and application development accounts for the highest proportion; domains such as language runtimes, data management, and data engineering are evenly distributed; while code related to platforms and security accounts for a lower proportion in the AI subset. The overall distribution is consistent with the actual investment and ecological scale of various software projects in real-world development environments \cite{OctoverseArchives2025,2024StackOverflow}, reflecting the dataset's good coverage of the in-the-wild software ecosystem.

Subsequently, we further analyzed the security characteristics of the dataset: using the CodeQL static analysis tool and an LLM-based judgment mechanism to scan all code samples, identifying CWE vulnerability scenarios and analyzing their distribution (as shown in Fig.\ref{fig:dataset_distributions}(\subref{fig:Dataset_CWE_Distribution})). The analysis results show that this evaluation dataset not only has broad application representativeness but also contains rich potential vulnerability features, serving as an effective benchmark for evaluating the security of AIGCode.

To further analyze the linguistic structural characteristics of code from different sources in the AIGCode evaluation set, this study proposes a lightweight Lexical Complexity Score (LCS). The LCS approximates the depth of logical branches in code by counting the number of control flow statements and logical operators. Its design references the ``decision point accumulation'' concept of McCabe's cyclomatic complexity, with a half-weight adjustment for logical operators introduced in the weight design to balance the contribution of semantic complexity. This metric can be efficiently calculated without relying on syntax parsing, making it suitable for large-scale code evaluation scenarios across multiple languages. Its definition is as follows:

\begin{equation} \label{eq:lcs}
\text{LCS} = 1 + N_{\text{cf}} + \frac{ N_{\text{op}}}{2}
\end{equation}

Here, $N_{\text{cf}}$ represents the number of occurrences of control flow statements and $N_{\text{op}}$ denotes the number of logical operators. For different programming languages, we adopt differentiated regular template matching rules for statistics to eliminate biases caused by differences in language syntax.

As shown in Fig.\ref{fig:dataset_distributions}(\subref{fig:Code Complexity Distribution}), the total dataset exhibits a significant bimodal difference in complexity distribution: most code falls within the LCS range of 0 to 20, while a small portion has an LCS exceeding 80. The LCS distributions of the AI-generated subset and the Human-written subset are roughly similar to that of the total dataset. Notably, the proportion of data with an LCS above 80 in the Human-written subset is significantly higher than in the AI-generated subset, which is consistent with the current situation where AI struggles to generate ultra-long code \cite{chen2021evaluating}.

\begin{table*}[htbp]
\centering
\caption{Comparative Distribution of Dataset Samples by Application Domain}
\label{tab:Application Domain}
\resizebox{\textwidth}{!}{%
\begin{tabular}{lrrrrrr}
\toprule
\multirow{2}{*}{\textbf{Application Domain}} &
\multicolumn{2}{c}{\textbf{All Dataset}} &
\multicolumn{2}{c}{\textbf{Human-written Subset}} &
\multicolumn{2}{c}{\textbf{AI-generated Subset}} \\
\cmidrule(lr){2-3} \cmidrule(lr){4-5} \cmidrule(lr){6-7}
 & \textbf{Count} & \textbf{ Proportion(\%)} & \textbf{Count} & \textbf{Proportion(\%)} & \textbf{Count} & \textbf{Proportion(\%)} \\
\midrule
Web and application development & 32\,835 & 41.21 & 10\,512 & 26.39 & 22\,323 & 56.04 \\
Language and runtime            &  9\,997 & 12.55 &  5\,489 & 13.78 &  4\,508 & 11.32 \\
Data management and persistence &  9\,676 & 12.14 &  4\,531 & 11.37 &  5\,145 & 12.92 \\
Data Science and Engineering    &  7\,584 &  9.52 &  5\,436 & 13.65 &  2\,148 &  5.39 \\
Network, Distribution and Security & 6\,023 &  7.56 &  3\,517 &  8.83 &  2\,506 &  6.29 \\
Operations and reliability      &  5\,748 &  7.21 &  3\,970 &  9.97 &  1\,778 &  4.46 \\
Client and Graphics             &  5\,512 &  6.92 &  4\,307 & 10.81 &  1\,205 &  3.02 \\
Platforms and Systems           &  1\,572 &  1.97 &  1\,406 &  3.53 &    166 &  0.42 \\
Others                          &    725 &  0.91 &    668 &  1.68 &     57 &  0.14 \\
\midrule
\textbf{Total} & \textbf{79\,672} & \textbf{100.0} & \textbf{39\,836} & \textbf{100.0} & \textbf{39\,836} & \textbf{100.0} \\
\bottomrule
\end{tabular}}
\end{table*}

\subsection{Benchmark Model Evaluation and Limitations Analysis}
To systematically evaluate the transferability of existing AIGC detection models in code detection tasks, this study selected seven representative AI-generated text detection models as benchmarks. These models are all based on the fundamental assumption that ``AI-generated content and human-generated content have identifiable differences in distribution characteristics,'' and have demonstrated good performance in natural language detection tasks. This study transfers them to the code detection scenario and verifies their effectiveness on the constructed AIGCode evaluation dataset.

To comprehensively evaluate the detection capabilities of the above models, we used four metrics: Accuracy, Precision, Recall, and F1-score. Accuracy characterizes the overall proportion of correctly classified instances. Precision quantifies the reliability of predictions labeled as “AI-generated,” whereas Recall measures the model’s ability to recover true AI-generated samples. F1-score provides a harmonic balance between Precision and Recall, offering a more stable indicator under class-imbalanced settings. The formal definitions of these metrics are provided in the Appendix.

However, experimental results show that when these models are transferred from text detection to code detection tasks, due to the fundamental differences between code and natural language in terms of grammatical structure and semantic expression, existing models face significant performance challenges. We randomly sampled 30\% of the samples from the evaluation dataset for testing, and the performance of the seven models is shown in Fig.\ref{fig:comprehensive_analysis}(\subref{fig:models_overview_and_performance}). The results show that existing models generally suffer from unbalanced detection capabilities, performing well on a single metric such as Precision or Recall, but failing to maintain high overall performance simultaneously. To address these limitations, this study proposes a cascaded-aggregation framework based on ensemble learning, which integrates the complementary advantages of multiple benchmark models through a hierarchical decision mechanism, effectively mitigating the biased detection problem of a single model, thereby improving the overall performance and balance of AIGCode detection.

\subsection{Cascade-Aggregation Framework}

Although traditional AIGC text detectors perform well on natural language tasks, they generally show an imbalance when transferred to code-detection tasks: precision and recall cannot be simultaneously maintained at high levels. This limitation primarily stems from inherent biases in a single model’s architecture or training data. To address this issue systematically, we propose a cascade-aggregation framework based on ensemble learning.

The core idea of this framework derives from ensemble strategies and aims to jointly leverage multiple base detectors with different decision tendencies. We first profile the performance of each base model on 30\% of the evaluation set (see Fig.\ref{fig:comprehensive_analysis}(\subref{fig:models_overview_and_performance})), and then categorize the models into two groups: high-precision and high-recall. High-precision models (e.g., {\ModelHCR}HCR) adopt conservative decision policies when labeling AIGCode and effectively suppress false positives; high-recall models (e.g., {\ModelORD}ORD) are more sensitive and strive to capture as many potential AI-generated samples as possible. Our framework exploits the complementarity of these two model types via a structured cascade, achieving precise decisions on high-confidence samples and strong recall in ambiguous regions—ultimately optimizing both ``high precision'' and ``high recall'' in tandem.

Based on this analysis, we exclude two underperforming models, {\ModelSRL}SRL and {\ModelYEnII}Y-En2, and use the remaining five more stable base models to build the cascade-aggregation framework. As illustrated in Fig.\ref{fig:Cascade-Aggregation Framework}, the framework consists of two core stages:

\begin{figure}[h]
\centering
\includegraphics[width=\columnwidth]{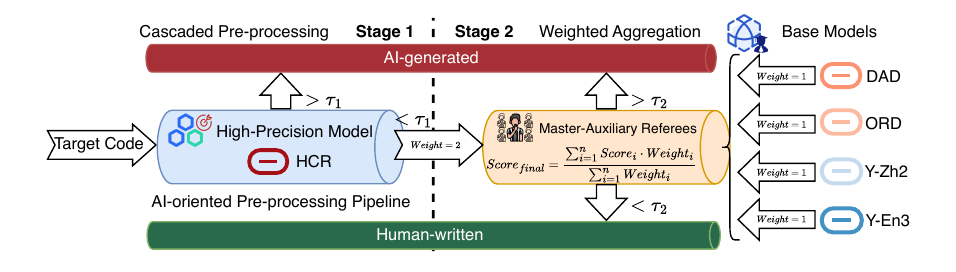}
\caption{Cascade-Aggregation Framework}
\label{fig:Cascade-Aggregation Framework}
\end{figure}

\textbf{Stage 1: Cascaded Preprocessing.} This stage forms an AI-oriented preprocessing pipeline inspired by early-exit mechanisms in ensemble learning. The pipeline uses the high-precision base model {\ModelHCR}HCR for fast screening: if its AI-generation confidence exceeds the preset threshold $\tau_{1}$ (set to 0.9), the sample is immediately labeled as ``AI-generated'', and the process terminates. This mechanism efficiently filters a portion of easily classified, high-confidence samples and significantly reduces the computational load for subsequent stages.

\textbf{Stage 2: Weighted Aggregation.} Samples not decided in Stage 1 proceed to a finer-grained evaluation. We adopt a master–auxiliary referee scheme for weighted aggregation: the highest-precision model {\ModelHCR}HCR is designated as the master referee with weight 2, while the other four base models ({\ModelYEnIII}Y-En3, {\ModelDAD}DAD, {\ModelYZhII}Y-Zh2, {\ModelORD}ORD) serve as auxiliary referees with weight 1 each. The final aggregated score $Score_{\text{final}}$ is computed as:

\begin{gather} \label{eq:score_final}
Score_{\text{final}} = \frac{\sum_{i=1}^{n} (Score_i \cdot Weight_i)}{\sum_{i=1}^{n} Weight_i}
\end{gather}

If $Score_{\text{final}}$ is below the decision threshold $\tau_{2}$ (set to 0.53), the sample is labeled ``human-written''; otherwise, it is labeled ``AI-generated.''

Experimental results (see Section~\ref{sec:evaluation-results}) show that the proposed cascade-aggregation framework significantly improves both precision and $F_{1}$-score. This validates that, for structured code-detection tasks, ensemble learning can overcome the performance bottlenecks of single models through model complementarity and information aggregation.

\subsubsection{Evaluation Results and Analysis Across Multiple Settings}
\label{sec:evaluation-results}

The above analysis reveals a precision–recall imbalance when traditional AIGC detectors are applied to code. To verify the effectiveness of our cascade-aggregation framework ({\ModelCAF}CAF), we conduct comprehensive comparative and ablation studies.

In the comparative study, we evaluate seven baseline models and {\ModelCAF}CAF on the held-out 70\% of the evaluation set, using Accuracy, Precision, Recall, and the harmonic mean $F_{1}$-score as metrics.

As shown in Fig.\ref{fig:comprehensive_analysis}(\subref{fig:Performance Comparison of CAF and Baseline Models}), {\ModelCAF}CAF achieves the best performance on three core metrics—Accuracy (0.718), Precision (0.716), and $F_{1}$-score (0.719). Although its Recall (0.722) is slightly lower than that of a few models specifically tuned for recall (e.g., {\ModelYZhII}Y-Zh2, {\ModelDAD}DAD, {\ModelORD}ORD), {\ModelCAF}CAF’s advantage is that it does not improve recall at the expense of precision. Instead, it attains a more favorable structural balance between the two, thereby yielding the best overall $F_{1}$-score. This indicates that {\ModelCAF}CAF effectively negotiates the trade-off between ``low false positives'' and ``low false negatives,'' rather than pursuing a single metric to the extreme.

This improvement does not come from simple majority voting but from the two-stage weighted decision mechanism. Stage 1 establishes a reliable decision boundary using the high-precision model, preventing human-written code from being misclassified; Stage 2 then applies master–auxiliary weighted aggregation to calibrate confidence on ambiguous samples, thereby maintaining high precision while robustly improving recall.

To further dissect {\ModelCAF}CAF, we design an ablation study comparing the full model with two variants: {\ModelStageI}\emph{w/o Stage 1} (removing cascaded preprocessing) and {\ModelStageII}\emph{w/o Stage 2} (removing weighted aggregation).

\begin{figure}[h]
    \centering
    \begin{subfigure}[t]{\textwidth}
        \centering
        \vspace{0pt}
        \providecommand{\drawModelPlots}{%
\begin{tikzpicture}[scale=0.6]

\pgfplotsset{
myaxis/.style={
ybar, bar width=12pt,
ymin=0, ymax=1.05,
ytick={0,0.25,0.5,0.75,1},
yticklabels={0,.25,.5,.75,1},
axis x line*=bottom, axis y line*=left,
axis line style={black!40},
tick style={black!50, line width=0.3pt},
ymajorgrids, grid style={black!10,dashed},
xtick=data,
xticklabel style={font=\normalsize, xshift=58pt,yshift=0pt},
ylabel style={font=\footnotesize},
title style={font=\bfseries\footnotesize},
width=6.75cm, height=3.8cm,
enlarge x limits=0.12,
nodes near coords,
point meta=rawy,
every node near coord/.style={
font=\scriptsize,
/pgf/number format/fixed,
/pgf/number format/precision=3,
color=black
}
}
}

\def\XTicks{    
\ModelHCR\hspace{4pt}
\ModelSRL\hspace{4pt}
\ModelORD\hspace{4pt}
\ModelDAD\hspace{4pt}
\ModelYZhII\hspace{4pt}
\ModelYEnII\hspace{4pt}
\ModelYEnIII
}

\begin{groupplot}[
group style={group size=4 by 1, horizontal sep=5mm, vertical sep=16mm},
myaxis,
xtick={1,2,3,4,5,6,7},
xticklabels=\XTicks
]
\nextgroupplot[title=Accuracy]
\addplot+[barHCR, bar shift=0pt] coordinates {(1,0.579)};
\addplot+[barSRL, bar shift=0pt] coordinates {(2,0.479)};
\addplot+[barORD, bar shift=0pt] coordinates {(3,0.5032)};
\addplot+[barDAD, bar shift=0pt] coordinates {(4,0.5118)};
\addplot+[barYZh, bar shift=0pt] coordinates {(5,0.5092)};
\addplot+[barYEn2, bar shift=0pt] coordinates {(6,0.4614)};
\addplot+[barYEn3, bar shift=0pt] coordinates {(7,0.5765)};

\nextgroupplot[title=Precision]
\addplot+[barHCR, bar shift=0pt] coordinates {(1,0.666)};
\addplot+[barSRL, bar shift=0pt] coordinates {(2,0.2191)};
\addplot+[barORD, bar shift=0pt] coordinates {(3,0.5016)};
\addplot+[barDAD, bar shift=0pt] coordinates {(4,0.5061)};
\addplot+[barYZh, bar shift=0pt] coordinates {(5,0.5045)};
\addplot+[barYEn2, bar shift=0pt] coordinates {(6,0.3811)};
\addplot+[barYEn3, bar shift=0pt] coordinates {(7,0.586)};

\nextgroupplot[title=Recall]
\addplot+[barHCR, bar shift=0pt] coordinates {(1,0.3164)};
\addplot+[barSRL, bar shift=0pt] coordinates {(2,0.0166)};
\addplot+[barORD, bar shift=0pt] coordinates {(3,0.9493)};
\addplot+[barDAD, bar shift=0pt] coordinates {(4,0.9662)};
\addplot+[barYZh, bar shift=0pt] coordinates {(5,0.9982)};
\addplot+[barYEn2, bar shift=0pt] coordinates {(6,0.1246)};
\addplot+[barYEn3, bar shift=0pt] coordinates {(7,0.5204)};

\nextgroupplot[title=F1 Score]
\addplot+[barHCR, bar shift=0pt] coordinates {(1,0.429)};
\addplot+[barSRL, bar shift=0pt] coordinates {(2,0.0308)};
\addplot+[barORD, bar shift=0pt] coordinates {(3,0.6564)};
\addplot+[barDAD, bar shift=0pt] coordinates {(4,0.6642)};
\addplot+[barYZh, bar shift=0pt] coordinates {(5,0.6703)};
\addplot+[barYEn2, bar shift=0pt] coordinates {(6,0.1878)};
\addplot+[barYEn3, bar shift=0pt] coordinates {(7,0.5512)};
\end{groupplot}

\end{tikzpicture}
}
        \drawModelPlots
        \caption{Overview and Performance of Baseline Detection Models}
        \label{fig:models_overview_and_performance}
        \vspace{-5pt}
    \end{subfigure}
    \hfill
    \vspace{5pt}
    \begin{subfigure}[t]{\textwidth}
        \centering
        \vspace{0pt}
        \providecommand{\drawModelPlots}{%
\begin{tikzpicture}[scale=0.6]

\pgfplotsset{
myaxis/.style={
ybar, bar width=12pt,
ymin=0, ymax=1.05,
ytick={0,0.25,0.5,0.75,1},
yticklabels={0,.25,.5,.75,1},
axis x line*=bottom, axis y line*=left,
axis line style={black!40},
tick style={black!50, line width=0.3pt},
ymajorgrids, grid style={black!10,dashed},
xtick=data,
xticklabel style={font=\normalsize, xshift=58pt,yshift=0pt},
ylabel style={font=\footnotesize},
title style={font=\bfseries\footnotesize},
width=6.75cm, height=3.8cm,
enlarge x limits=0.12,
nodes near coords,
point meta=rawy,
every node near coord/.style={
font=\scriptsize,
/pgf/number format/fixed,
/pgf/number format/precision=3,
color=black
}
}
}

\def\XTicks{    
\ModelHCR\hspace{8pt}
\ModelORD\hspace{8pt}
\ModelDAD\hspace{8pt}
\ModelYZhII\hspace{8pt}
\ModelYEnIII\hspace{8pt}
\ModelCAF
}

\begin{groupplot}[
group style={group size=4 by 1, horizontal sep=5mm, vertical sep=16mm},
myaxis,
xtick={1,2,3,4,5,6,7},
xticklabels=\XTicks
]
\nextgroupplot[title=Accuracy]
\addplot+[barHCR, bar shift=0pt] coordinates {(1,0.5814)};
\addplot+[barORD, bar shift=0pt] coordinates {(2,0.5049)};
\addplot+[barDAD, bar shift=0pt] coordinates {(3,0.5108)};
\addplot+[barYZh, bar shift=0pt] coordinates {(4,0.5094)};
\addplot+[barYEn3, bar shift=0pt] coordinates {(5,0.5764)};
\addplot+[barCAF, bar shift=0pt] coordinates {(6,0.7181)};

\nextgroupplot[title=Precision]
\addplot+[barHCR, bar shift=0pt] coordinates {(1,0.6702)};
\addplot+[barORD, bar shift=0pt] coordinates {(2,0.5027)};
\addplot+[barDAD, bar shift=0pt] coordinates {(3,0.5058)};
\addplot+[barYZh, bar shift=0pt] coordinates {(4,0.5049)};
\addplot+[barYEn3, bar shift=0pt] coordinates {(5,0.5862)};
\addplot+[barCAF, bar shift=0pt] coordinates {(6,0.7164)};

\nextgroupplot[title=Recall]
\addplot+[barHCR, bar shift=0pt] coordinates {(1,0.3213)};
\addplot+[barORD, bar shift=0pt] coordinates {(2,0.952)};
\addplot+[barDAD, bar shift=0pt] coordinates {(3,0.9679)};
\addplot+[barYZh, bar shift=0pt] coordinates {(4,0.9987)};
\addplot+[barYEn3, bar shift=0pt] coordinates {(5,0.5208)};
\addplot+[barCAF, bar shift=0pt] coordinates {(6,0.7223)};

\nextgroupplot[title=F1 Score]
\addplot+[barHCR, bar shift=0pt] coordinates {(1,0.4344)};
\addplot+[barORD, bar shift=0pt] coordinates {(2,0.6579)};
\addplot+[barDAD, bar shift=0pt] coordinates {(3,0.6644)};
\addplot+[barYZh, bar shift=0pt] coordinates {(4,0.6707)};
\addplot+[barYEn3, bar shift=0pt] coordinates {(5,0.5516)};
\addplot+[barCAF, bar shift=0pt] coordinates {(6,0.7194)};
\end{groupplot}

\end{tikzpicture}
}
        \drawModelPlots
        \caption{Performance Comparison of CAF and Baseline Models}
        \label{fig:Performance Comparison of CAF and Baseline Models}
        \vspace{-5pt}
    \end{subfigure}    
    \hfill
    \vspace{5pt}
    \begin{subfigure}[t]{0.45\textwidth}
    \centering
    \vspace{5pt}
    \providecommand{\drawModelPlots}{%
\begin{tikzpicture}[scale=1]

\pgfplotsset{
myaxis/.style={
ybar, bar width=8pt,
xmin=0.5, xmax=3.5,
ymin=0, ymax=1.05,
ytick={0,0.25,0.5,0.75,1},
yticklabels={0,.25,.5,.75,1},
axis x line*=bottom, axis y line*=left,
axis line style={black!40},
tick style={black!50, line width=0.3pt},
ymajorgrids, grid style={black!10,dashed},
xtick=data,
xticklabel style={font=\normalsize, xshift=15.5pt,yshift=0pt},
ylabel style={font=\footnotesize},
title style={font=\bfseries\footnotesize},
width=3.8cm, height=2.75cm,
enlarge x limits=0.12,
nodes near coords,
point meta=rawy,
every node near coord/.style={
font=\scriptsize,
/pgf/number format/fixed,
/pgf/number format/precision=3,
color=black
}
}
}

\def\XTicks{    
\ModelCAF\hspace{1.5pt}
\ModelStageI\hspace{1.5pt}
\ModelStageII
}

\begin{groupplot}[
group style={group size=2 by 2, horizontal sep=5mm, vertical sep=13mm},
myaxis,
xtick={1,2,3,4,5,6,7},
xticklabels=\XTicks
]
\nextgroupplot[title=Accuracy]
\addplot+[barCAF, bar shift=0pt] coordinates {(1,0.7181)};
\addplot+[barStage1, bar shift=0pt] coordinates {(2,0.5057)};
\addplot+[barStage2, bar shift=0pt] coordinates {(3,0.4895)};

\nextgroupplot[title=Precision]
\addplot+[barCAF, bar shift=0pt] coordinates {(1,0.7164)};
\addplot+[barStage1, bar shift=0pt] coordinates {(2,0.5029)};
\addplot+[barStage2, bar shift=0pt] coordinates {(3,0.1868)};

\nextgroupplot[title=Recall]
\addplot+[barCAF, bar shift=0pt] coordinates {(1,0.7223)};
\addplot+[barStage1, bar shift=0pt] coordinates {(2,0.9973)};
\addplot+[barStage2, bar shift=0pt] coordinates {(3,0.0061)};

\nextgroupplot[title=F1 Score]
\addplot+[barCAF, bar shift=0pt] coordinates {(1,0.7194)};
\addplot+[barStage1, bar shift=0pt] coordinates {(2,0.6687)};
\addplot+[barStage2, bar shift=0pt] coordinates {(3,0.0119)};

\end{groupplot}

\end{tikzpicture}
}
    \drawModelPlots
    \captionsetup{justification=centering}
    \caption{Ablation Study: Impact of Components on CAF Performance}
    \label{fig:Ablation Study - Impact of Components on CAF Performance}
    \end{subfigure}
    \hfill%
    \begin{subfigure}[t]{0.45\textwidth}
    \centering
    \vspace{10pt}
    \setlength{\tabcolsep}{2.5pt}
    \footnotesize
    \scalebox{1.1} {
        \begin{tabular}{lcc}
        \toprule
        \textbf{Model.Id} & \textbf{Source} & \textbf{HF downloads} \\
        \midrule
        \ModelHCR~HCR & arXiv [cs.CL] & 12,454 \\
        \ModelSRL~SRL & Huggingface & 60,717 \\
        \ModelORD~ORD & OpenAI Report & 71,494 \\
        \ModelDAD~DAD & Huggingface & 72 \\
        \ModelYZhII~Y-Zh2 & ICLR 2024 [cs.CL] & 377 \\
        \ModelYEnII~Y-En2 & ICLR 2024 [cs.CL] & 50 \\
        \ModelYEnIII~Y-En3 & ICLR 2024 [cs.CL] & 142 \\
        \ModelCAF~CAF & / & / \\
        \ModelStageI~CAF & w/o Stage1 & / \\
        \ModelStageII~CAF & w/o Stage2 & / \\
        \bottomrule
        \end{tabular}
    }
    \vspace{5pt}
    \caption{Symbols and Details of Models}
    \label{fig:model_details}
\end{subfigure}    
    \caption{Comprehensive analysis of CAF and baseline models}
    \label{fig:comprehensive_analysis}
\end{figure}

As is shown in Fig.\ref{fig:comprehensive_analysis}(\subref{fig:Ablation Study - Impact of Components on CAF Performance}), removing Stage 1 causes Accuracy and Precision to drop sharply to about 0.506, while Recall surges to 0.997. Without the initial high-confidence filter, the model becomes overly sensitive—able to detect nearly all AI samples but at the cost of many false positives—resulting in an $F_{1}$-score of 0.669 and reduced reliability.

Removing Stage 2 leads to catastrophic failure, with performance collapsing ($F_{1}$-score $= 0.012$). This demonstrates that weighted aggregation is critical for forming an effective decision boundary and for preserving basic discriminative capability.

Overall, the evaluation corroborates {\ModelCAF}CAF’s effectiveness from two angles: the comparative study shows that {\ModelCAF}CAF outperforms existing baselines in overall discrimination, while the ablation study confirms that both stages are indispensable—especially the decisive role of Stage 2. Our cascade-aggregation framework provides a practical and effective pathway to resolve the precision–recall imbalance in AIGCode detection.

\section{Collection and Construction of In-the-Wild Code Data}

This study relies on two primary categories of code data: open-source community code and code-vulnerability threat intelligence. The former is collected by systematically harvesting project evolution histories on GitHub to broadly reflect the real ecology of modern software development; the latter is used to precisely capture recently disclosed vulnerability threats and their remediation patterns, providing support for security-critical analyses.

\paragraph{Open-Source Code Dataset: Construction Pipeline and Distribution Characteristics}
To construct a large-scale dataset of code changes representative of real development activity, we design a systematic data collection strategy centered on the GitHub platform.

The core of collection is selecting projects with high activity and strong community impact. We use the number of Stars as a key proxy for project popularity and quality and select the top 1,000 open-source repositories \cite{borges2016predicting}. The time window is set from January 1, 2022 to mid-2025 to cover recent development practices. The granularity is refined to the file level: for each repository within this window, we iterate over every commit, enumerate all files whose contents changed, and extract both the pre-change and post-change versions. Each such ``code pair'' constitutes a basic analytical unit in the dataset\cite{li2024understanding} . This process yields a large-scale, multilingual dataset of code evolution derived from genuine development activity, forming the foundation for subsequent analyses.

Commit activity continuously covers the entire target window from January 2022 to mid-2025, with density that fluctuates yet remains overall stable, showing a slight upward trend. This long-horizon, high-coverage profile ensures strong representativeness along the time dimension and faithfully reflects continuous integration and evolutionary patterns in software development.

\paragraph{Vulnerability Code Dataset: Acquisition Channels and Severity Analysis}

To support in-depth, empirical vulnerability analysis, we systematically construct a high-quality CVE vulnerability code dataset. Built entirely from public intelligence sources, the pipeline proceeds as follows: first, we retrieve authoritative vulnerability metadata from the official CVE website; next, we search open-source platforms such as GitHub using CVE identifiers and vulnerability-related keywords; finally, we perform manual verification and de-noising on the retrieved code snippets to ensure that each sample contains a paired vulnerable fragment and its patched version, accompanied by structured metadata such as the CVE identifier and vulnerability category.

\begin{figure}[htbp]
 
    \begin{subfigure}[b]{0.355\linewidth}
        \centering
        \includegraphics[width=\linewidth]{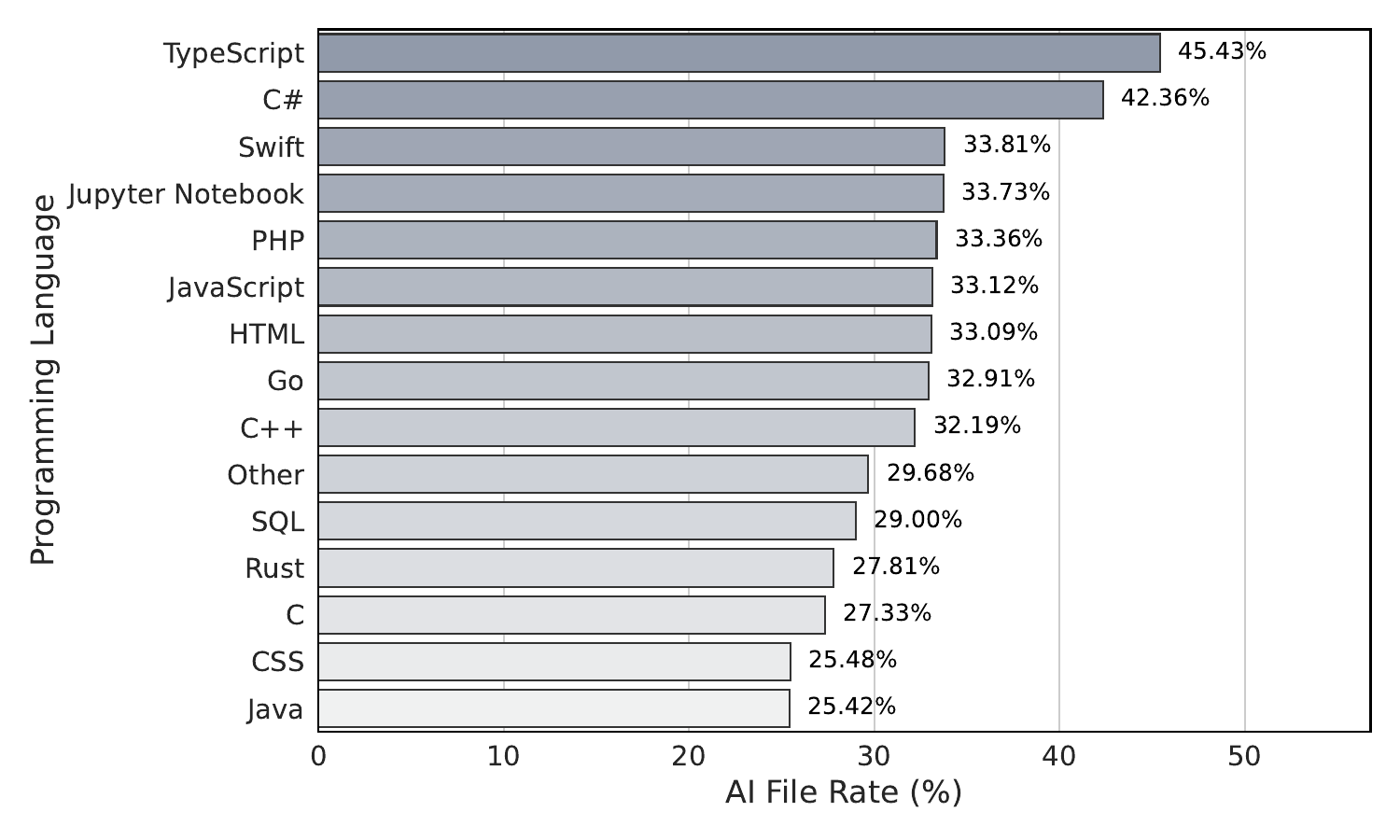}  
        \caption{AI Adoption by Language}
        \label{fig:5-1-ai file rate by language}
    \end{subfigure}
    \begin{subfigure}[b]{0.635\linewidth}
        \centering
        \includegraphics[width=\linewidth]{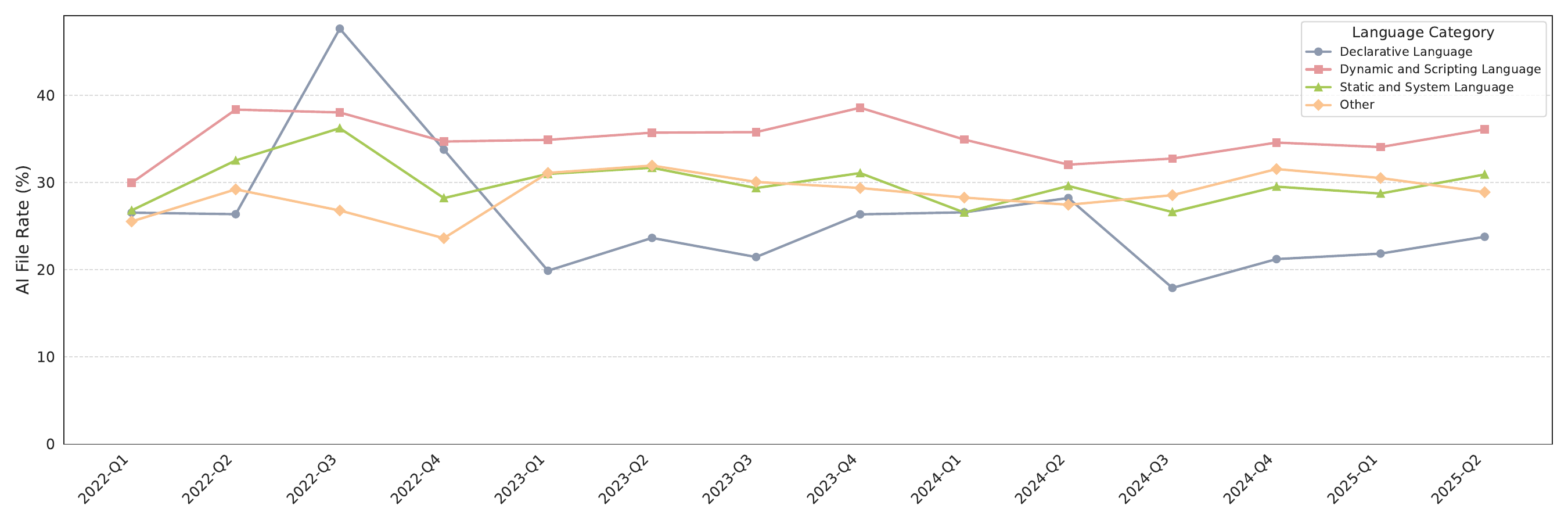} 
        \caption{AI Adoption by Tech Stack}
        \label{fig:5-2-ai rate by 4 categories}
    \end{subfigure}

    \begin{subfigure}[t]{0.3\linewidth}
        \centering
        \includegraphics[width=\linewidth]{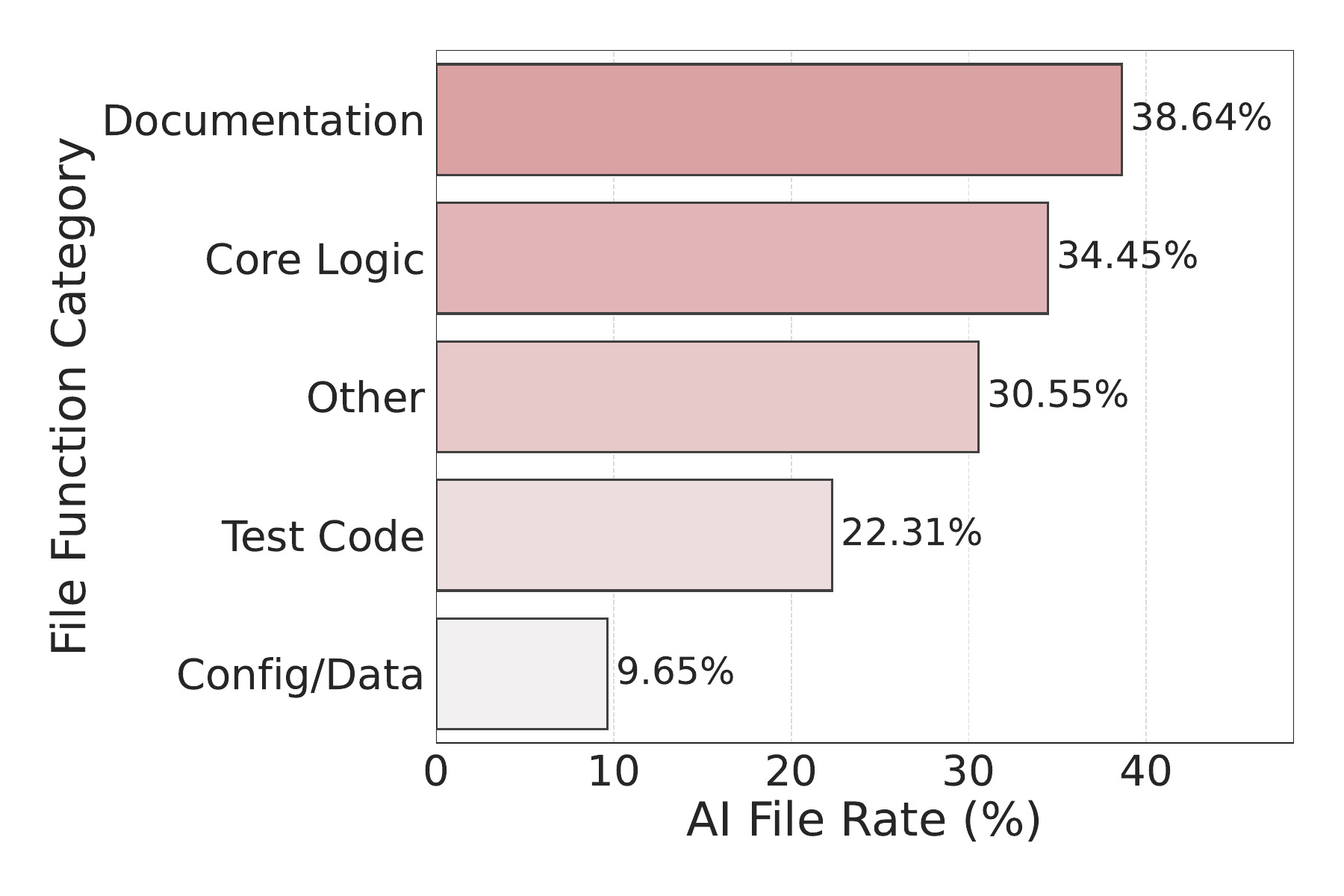} 
        \caption{AI Adoption by Code Function}
        \label{fig:5-3-ai rate by func}
    \end{subfigure}
    \begin{subfigure}[t]{0.69\linewidth}
        \centering
        \includegraphics[width=\linewidth]{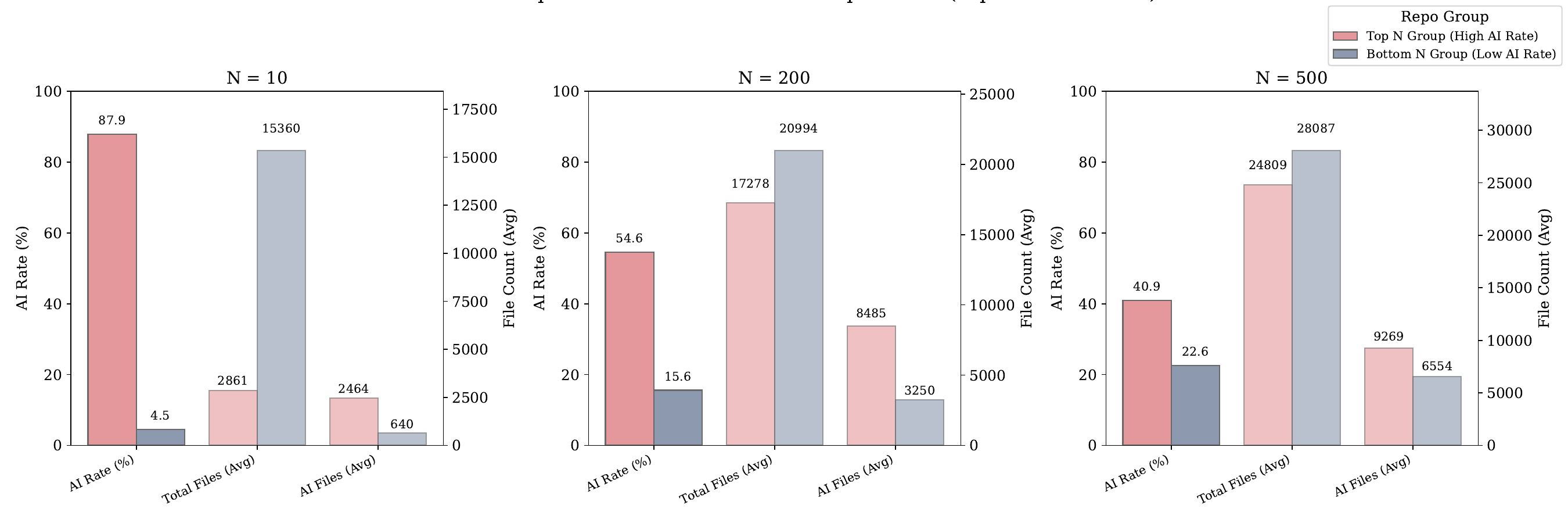} 
        \caption{Project Scale and AI Rate Group Comparison}
        \label{fig:5-4-repo extreme analysis}
    \end{subfigure}
    \centering
    \caption{Multidimensional Empirical Analysis of AI Code Generation in Wild Code}
    \label{fig:5-analysis of ai code}
\end{figure}

\section{An Empirical Study on the Adoption and Ecosystem of AIGCode}

\subsection{Adoption Landscape Across Mainstream Programming Languages}

Analysis of the dataset used in this study reveals significant differences in the adoption of AIGCode across mainstream programming languages, thereby highlighting the varying penetration and acceptance of AI-assisted programming technology in different technological ecosystems. We analyzed commit data from the top 1000 GitHub repositories between 2022 and 2025 to calculate the ``AI file rate'' for each language—the percentage of files in the final code version identified as AIGCode. The statistical results, shown in Fig.\ref{fig:5-analysis of ai code}(\subref{fig:5-1-ai file rate by language}), indicate that the AI file rate for different languages ranges from 25.42\% to 45.43\%, confirming the profound impact of language characteristics on the applicability and adoption of AIGCode.

\textbf{First, highly structured and strongly-typed languages demonstrate the highest AIGCode adoption rates. }TypeScript (45.43\%) and C\# (45.36\%) rank highest, suggesting that when AI models process code that is well-structured, type-constrained, and consistent with modern paradigms, they produce output that is of higher quality, less ambiguous, and ultimately more acceptable to developers, leading to increased practical adoption. This finding emphatically confirms the significant advantage of AI-assisted programming tools in generating highly predictable, structured code.

\textbf{Secondly, the Web technology stack and general-purpose scripting languages have formed a stable and widespread AIGCode-assisted model.} The AI document rate for languages such as JavaScript, PHP, HTML, and Jupyter Notebook remains stable at a mid-to-high level of around 33\%. This concentrated distribution indicates that in high-iteration-speed scenarios such as Web front-end and back-end development, data processing, and real-time documentation, AIGCode has become a stable tool providing continuous efficiency gains and is deeply integrated into standard development processes. This proves that AIGCode assistance is no longer limited to generating boilerplate code or simple scripts, but effectively supports the implementation of actual business logic.

Finally, developers exhibit pronounced caution regarding AIGCode adoption in domains demanding extremely high system stability and stringent resource control. Correspondingly, the adoption rates for enterprise and system-level languages, such as Java (25.42\%) and C (27.33\%), are significantly lower across our statistical range. This outcome aligns with their prevalence in large, complex production environments characterized by demanding stability requirements.This conservative tendency likely stems from two key factors: First, the complex contextual demands of extensive legacy systems challenge the ability of AI to generate accurate and relevant code. Second, the high cost associated with code errors in these critical domains necessitates elevated verification standards for AIGCode outputs and their applicability. Consequently, the widespread adoption of AIGCode in such highly standardized environments is fundamentally constrained by the imperative for continuous, rigorous verification of the security and accuracy of the generated code.

\subsection{Temporal Evolution of Adoption Rates Across Technology Stacks}

To reveal the penetration paths and maturity differences of AIGCode across different programming paradigms, we conducted a time-series analysis of AIGC document rates across existing data categorized into four language types: dynamic and scripting languages, static and system languages, declarative languages, and others. The study shown in Fig.\ref{fig:5-analysis of ai code}(\subref{fig:5-2-ai rate by 4 categories}) found that the adoption trajectories of each category are highly correlated with the accuracy, flexibility, and reliability required by their core paradigms, clearly outlining the evolution of AIGC technology from experimental assistance to production-grade tools.

\textbf{The declarative language category exhibits the highest initial adoption fluctuation, reflecting the experimental nature and limitations of AIGC in structured text generation. }As a paradigm encompassing configuration, document markup, and data querying, its AIGC document rate rapidly reached an early peak in Q3 2022 before significantly declining and remaining stable below 30\% for an extended period. This phenomenon indicates that although AI can efficiently generate patterned declarative content, its results are highly dependent on project-specific contexts, posing challenges in accuracy and maintainability. Consequently, after initial exploration, its adoption rate has stabilized at a relatively conservative level.

\textbf{The Dynamic and Scripting Languages category exhibits a consistently high adoption rate, solidifying the role of AIGCode as a core tool in rapid iterative development.} This category, which includes languages such as Python and JavaScript, has maintained its position among the top tiers in AIGCode usage rates since the beginning of our observation period, approaching 38\% again by Q2 2025. This steady upward trend proves that the efficiency gains provided by AIGC in typical scenarios such as rapid prototyping, automated scripting, and business logic implementation have been widely recognized by developers and have been deeply integrated into the development workflow as a reliable and frequently used auxiliary tool.

\textbf{The adoption trajectory of the static and system language categories is the most revealing, with fluctuations directly reflecting developers' stringent requirements for code quality and tool maturity.} The AIGC file rate in this category experienced a significant pullback from its peak in Q3 2022, hitting a low of approximately 19\% in Q1 2023, before gradually recovering and stabilizing at around 30\%. This trend indicates that in paradigms like C/C++, Rust, and Java, used for building high-performance, high-reliability systems, developers are extremely cautious about introducing AIGC due to the high cost of code errors. The early decline revealed the model's inadequacy in handling complex memory management, performance optimization, and deep abstractions, while the subsequent steady recovery signifies that improvements in model capabilities and AI code tools have gradually won over developers' trust.

\subsection{Distribution Characteristics Across File Functions}

To quantify the penetration level of AIGCode at different stages of the software development lifecycle, we categorized it into five functional groups based on file path and extension: Documentation, Core Logic, Test Code, Config/Data, and Other. These functional codes typically appear in clusters at different development stages, and we calculated the average AI document rate for each functional category. Statistical results presented in Fig.~\ref{fig:5-analysis of ai code}(\subref{fig:5-3-ai rate by func}) further show that the adoption rate of AIGCode varies significantly across different functional categories, with the highest value (Documentation, 38.64\%) differing from the lowest value (Config/Data, 9.65\%) by approximately 30 percentage points, clearly revealing the task preferences and application boundaries of current AI-assisted programming tools in actual development.

\textbf{First, AIGCode exhibits a significantly higher adoption rate in documentation generation tasks compared to other stages, underscoring its high application value and relatively low risk profile in structured text generation.} The Documentation category ranks first with an AI document rate of 38.64\%. This is mainly because document content usually has strong patterned characteristics and a higher error tolerance than executable code. Therefore, developers are more inclined to use AI to automate this type of repetitive text work with lower creative requirements.

\textbf{Secondly, AIGCode has penetrated deep into the core business logic construction stage, demonstrating its considerable auxiliary capabilities in functional implementation tasks.} The AI file rate in the Core Logic category reached 34.45\%, significantly higher than that of test and configuration files. This result indicates that AIGCode is no longer limited to peripheral or auxiliary tasks, but has been accepted by developers and used to implement critical business functions, reflecting a considerable degree of recognition for its generated quality in terms of logical correctness and syntactic compliance.

\textbf{Meanwhile, developers are showing a more cautious approach to test code generation, indicating higher requirements for the reliability of AI in quality assurance.} The AI file rate for the Test Code category is 22.31\%, significantly lower than Core Logic. Test code must rigorously cover various boundary conditions and abnormal paths; its correctness directly affects software quality. Therefore, developers have higher verification standards for the accuracy of AI-generated test cases, and human intervention and review remain indispensable in this stage.

\textbf{Finally, the extremely low AI file rate (9.65\%) in the Config/Data category reveals the fundamental limitations of AIGCode in handling tasks with high context dependencies and strong formatting constraints.} Even minor deviations in configuration files and data scripts can lead to system-level failures; therefore, developers generally prefer to ensure their correctness through manual writing and rigorous review. In tasks with such high accuracy requirements, the penetration of AIGCode is significantly suppressed.

\subsection{Correlation between Repository Scale and Adoption Patterns}

To explore the relationship between project characteristics and AIGC adoption rates, we systematically compared the differences in average project size and absolute AIGC usage between repositories with the highest (Top N) and lowest (Bottom N) adoption rates, based on four grouping granularities: N=10, 100 and 500. The analytical metrics included average AI document rate, average total number of documents, and average number of AIGC documents. The results in Fig.\ref{fig:5-analysis of ai code}(\subref{fig:5-4-repo extreme analysis}) show that although the characteristic values of both groups systematically change with increasing N, the relative difference pattern remains stable, revealing a unique relationship between AIGC adoption and project size.

\textbf{First, high AIGC adoption rates are significantly concentrated in small to medium-sized repositories, suggesting their tool positioning in specific scenarios.} In the most stringent grouping of $N=10$, the average total number of files in the Bottom $N$ group (15,359.7) is approximately six times that of the Top $N$ group (2,660.8); even when $N$ increases to 500, the size difference remains significant. This strong correlation between size characteristics and high AI file rates indicates that AIGC is more likely to achieve deep penetration in projects with smaller codebases, lower context loads, or highly specialized tasks at the current stage, while its overall adoption is limited in large repositories with complex structures and strict specifications.

\textbf{Secondly, the absolute usage of AIGC in low-adoption-rate repositories remains significant, highlighting its auxiliary value in large-scale scenarios. }Although the AI file rate in the Bottom $N$ group is significantly lower than that in the Top $N$ group, the absolute number of AIGC files remains considerable with project size. For example, when $N=10$, the average number of AIGC files in the Top $N$ group (2,464.3) is approximately four times that of the Bottom $N$ group (640.5); however, when $N$ increases to 500, this gap narrows significantly (9,268.6 vs. 6,553.6). This indicates that although the file-level penetration rate is low in large projects, AIGC still undertakes a considerable amount of auxiliary coding work in actual development due to its large codebase.

In summary, AIGC adoption patterns are not simply linearly related to project size, but rather reflect its differentiated role across different engineering stages and project types. High adoption rates are primarily seen in small to medium-sized, high-iteration, or specialized projects, where AIGC plays a core supporting role. In large, mature repositories, however, AIGC is used cautiously in a ``low proportion, high total volume'' manner, its penetration depth constrained by both existing engineering standards and code complexity.

\section{An Empirical Analysis of Security Risks in AIGCode}

\begin{figure}[htbp]
 
    \begin{subfigure}{\linewidth} 
        \centering
        \includegraphics[width=0.97\linewidth]{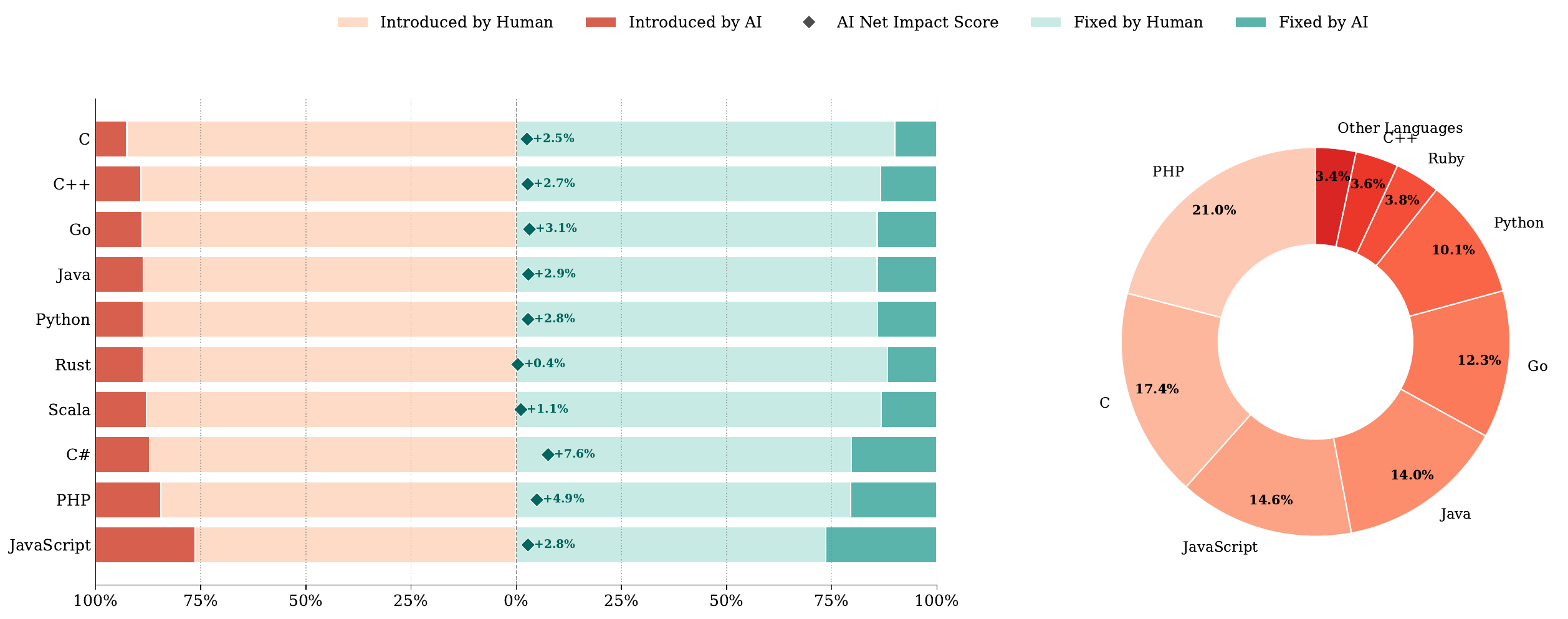}
        \caption{AI Role and Net Impact in Vulnerability Lifecycle}
        \label{fig:6-1-roles in the life circle}
    \end{subfigure}
    \begin{subfigure}{\linewidth} 
        \centering
        \includegraphics[width=0.95\linewidth]{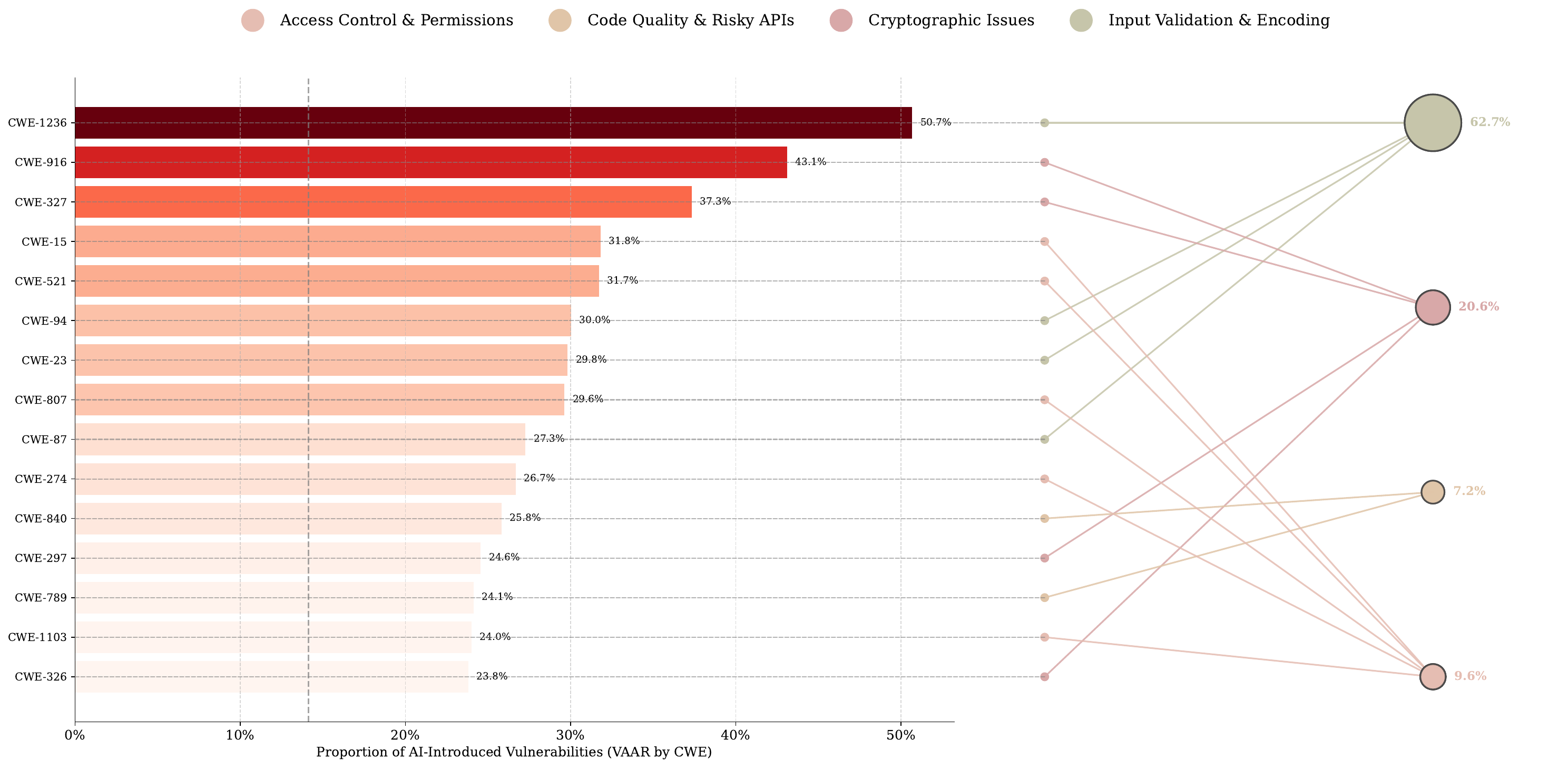}
        \caption{AI's CWE Preference and Risk Category Mapping}
        \label{fig:6-2-cwe category}
    \end{subfigure}
    \begin{subfigure}[t]{0.35\linewidth}
        \centering
        \includegraphics[width=\linewidth]{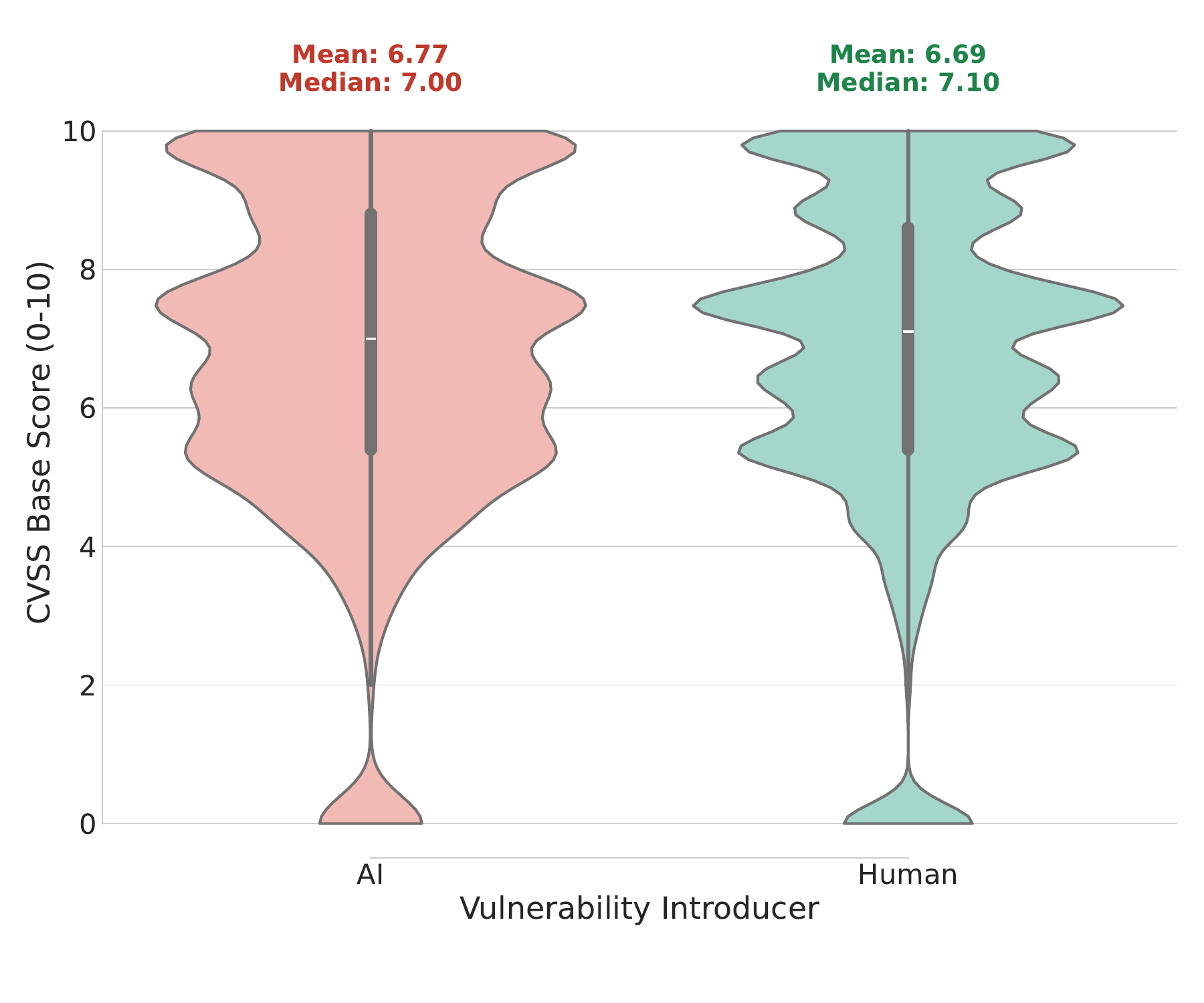} 
        \caption{Severity Distribution}
        \label{fig:6-3-cvss}
    \end{subfigure}
    \begin{subfigure}[t]{0.55\linewidth}
        \centering
        \includegraphics[width=\linewidth]{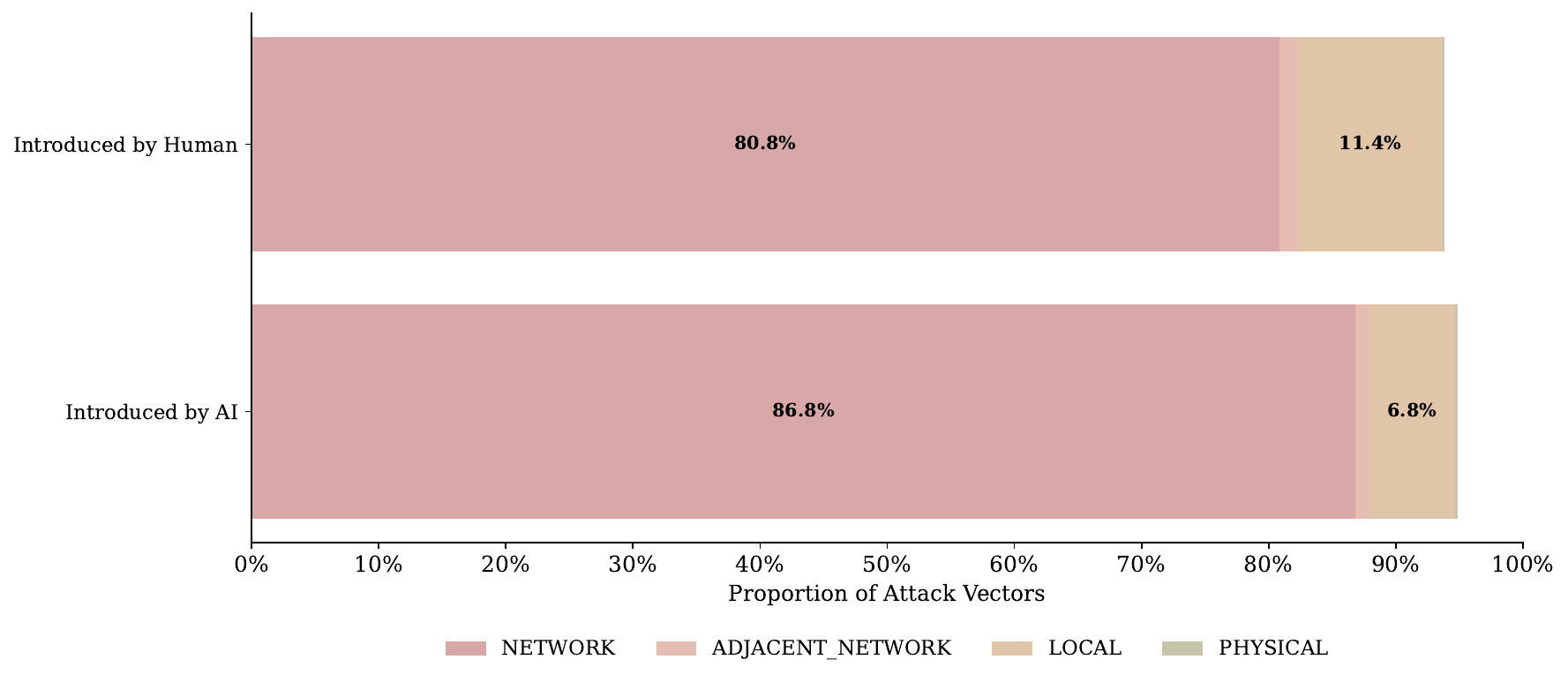} 
        \caption{Attack Vector Characteristics of AI-Introduced }
        \label{fig:6-4-attack vector}
    \end{subfigure}
    \centering
    \caption{Empirical Analysis of AI Code Generation Impact on Code Security: Roles, Types, and Severity}
    \label{fig:6-analysis of sec}
\end{figure}

To systematically assess the security risks of AIGCode, this study conducts an empirical analysis from three levels: at the macro level, comparing the differences in roles between AI and humans in the vulnerability lifecycle; at the micro level, exploring the type preferences of vulnerabilities introduced by AI; and at the comprehensive level, assessing the severity and risk characteristics of vulnerabilities introduced by AI, thereby constructing a complete security risk profile of AIGC.

\subsection{The Dual Role in Vulnerability Lifecycle: Introduction and Remediation}

To quantify the roles of AI as an introducer and fixer in the software vulnerability lifecycle, and to evaluate its net impact on code security, we analyzed the proportion of vulnerability introduction sources (human vs. AI) and fix sources (human vs. AI) across various programming languages. We then calculated the AI net impact score, defined as the AI introduction proportion minus the AI fix proportion. The horizontal stacked bar chart in Fig.\ref{fig:6-analysis of sec}(\subref{fig:6-1-roles in the life circle}) illustrates the distribution of AIGCode's impact on security, while the pie chart on the right shows the proportion of each language in the entire vulnerability sample. Overall observations indicate that the AI net impact score is positive across all analyzed languages—meaning the AI introduction proportion is greater than the AI fix proportion. This initially suggests that AI currently contributes more to vulnerability introduction than to vulnerability remediation.

\textbf{The proportion of AI-introduced vulnerabilities is lower than that of human-introduced ones across all languages, but AI's contribution to vulnerability introduction is unusually high in certain languages}—indicating that AI poses greater security risks in high-redundancy or pattern-based programming scenarios. Although human-introduced vulnerabilities dominate the total vulnerability sample (exceeding 75\%), AI exhibits relatively higher introduction proportions in C\# (AI net impact score: +7.6\%) and PHP (AI net impact score: +4.9\%). These two languages are often used in web development and application layers, involving numerous API calls and I/O operations with high degrees of patternization and repetitiveness. The higher AI introduction proportion suggests that when AI attempts to rapidly generate such pattern-based code, it may be more prone to introducing flaws—especially in scenarios involving security boundaries or complex configurations.

\textbf{In the current vulnerability lifecycle, AIGCode's net impact on security is relatively positive}—the AI net impact score is positive across all analyzed languages, meaning AI's risk of fixing vulnerabilities is consistently higher than its contribution to introducing them. Whether for languages with high security requirements such as Rust (+0.4\%) and Scala (+1.1\%), or application-layer languages like PHP (+4.9\%) and C\# (+7.6\%), the AI net impact score falls to the right of the central axis. Although scores for languages like Rust are extremely low—indicating a near-neutral impact on security—the overall trend still confirms that AIGCode's net contribution to security issues is negative. This underscores that when relying on AIGCode for programming, developers must regard manual security audits as an indispensable step to mitigate the additional security burden brought by AI.

\textbf{The closer the technology stack is to the underlying layer, the smaller the marginal effect of AI introducing security vulnerabilities}. For system-level and low-level control languages, the AI net impact score is significantly lower than that of application-layer languages. Rust has an AI net impact score of only +0.4\%, the lowest among all languages, while the scores for C and C++ also remain at relatively low levels of +2.5\% and +2.7\%, respectively. This trend may stem from two aspects: first, developers working with low-level languages are more cautious when integrating AIGCode and conduct stricter security verification; second, when AI models generate code involving complex low-level logic such as memory management and pointer operations, their uncertainty leads to a inherently low adoption rate in security-critical paths, thereby reducing the absolute proportion and net impact of vulnerabilities introduced by AI.

\subsection{Taxonomy and Patterns of AI-Introduced Vulnerabilities}

To identify the CWE types preferred by AIGCode when introducing security vulnerabilities, map them to higher-level security risk categories, and thereby provide targeted guidance for mitigation measures, we calculated the proportion of vulnerabilities introduced by AI in each specific CWE type. We then used the mapping diagram on the right to group these CWEs into predefined risk categories. The horizontal bar chart in Fig.\ref{fig:6-analysis of sec}(\subref{fig:6-2-cwe category}) illustrates the concentration of vulnerability types introduced by AI, with the AI file rate of the top ranked CWE types far exceeding the average level.

\textbf{AIGCode exhibits significant type concentration in vulnerability introduction, showing a strong preference for specific CWEs related to unsafe operations and input handling}. Statistical results indicate that the two CWE types with the highest proportion of AI-introduced vulnerabilities are CWE-1236 (Use of Incorrectly Escaped Output) and CWE-916 (Use of a Broken or Risky Cryptographic Algorithm), with their AI file rate reaching 50.7\% and 43.1\% respectively. This means that over 40\% of all samples with these two types of vulnerabilities are caused by AI code generation. Such high concentration suggests that AIGCode models may systematically introduce flaws when handling API calls involving encoding, escaping, secure hashing, or cryptographic functions, likely due to a lack of in-depth understanding of security contexts and API best practices.

\textbf{Flaws in AIGCode stem from pattern recognition rather than high-level logical errors}. Vulnerabilities introduced by AI are mainly concentrated in the two major risk categories of ``Input Validation and Encoding'' and ``Code Quality and Risky APIs''. The right panel of Fig.\ref{fig:6-analysis of sec}(\subref{fig:6-2-cwe category}) shows that most of the top ranked CWEs are closely associated with either ``Input Validation and Encoding'' (accounting for approximately 62.7\%) or ``Code Quality and Risky APIs'' (accounting for approximately 20.6\%). This further confirms that the security flaws of AIGCode do not stem from complex business logic vulnerabilities or permission bypasses (as ``Access Control and Permissions'' only accounts for 9.6\%), but rather from unsafe API usage at the code snippet level, incorrect input sanitization, or improper encoding or escaping handling—all of which are common blind spots in code pattern recognition.

\subsection{Severity Assessment and Risk Profiling of Vulnerabilities}

Based on the analysis confirming AI's preference for introducing security vulnerabilities, we further analyzed the severity and risk characteristics of these vulnerabilities and compared them with vulnerabilities introduced by humans. The analysis is based on two core dimensions of the Common Vulnerability Scoring System (CVSS):

We used the Mann-Whitney U test to compare the severity score distributions of vulnerabilities introduced by AI and humans, setting the significance level at $\alpha$ = 0.05. The null hypothesis $H_0$ was that there was no difference in the medians between the two groups, and the alternative hypothesis $H_1$ was that there was a difference. The analysis revealed that although there was no significant difference in the macro-level severity of vulnerabilities introduced by AI and humans, there were significant differences in their micro-level risk characteristics.

As shown in Fig.\ref{fig:6-analysis of sec}(\subref{fig:6-3-cvss}), the Mann-Whitney U test results ($p=0.091$) indicate that there is no statistically significant difference in the severity distribution of vulnerabilities introduced by AI and humans. Descriptive statistics further support this conclusion: the median severity of vulnerabilities introduced by AI (7.00) is very close to that of humans (7.10), and the means (AI: 6.77, Human: 6.69) are also not significantly different.

However, in the attack vector analysis shown in Fig.\ref{fig:6-analysis of sec}(\subref{fig:6-4-attack vector}), we found that AI-introduced vulnerabilities accounted for a staggering 86.8\% of network attack vectors, significantly higher than the 80.8\% for human vulnerabilities. This indicates that flaws in AIGCode are more easily exposed to remote attack surfaces, resulting in a wider range of potential harm. Correspondingly, AI was more conservative in introducing vulnerabilities requiring local privileges (6.8\% vs. 11.4\%).

Based on the above analysis, we have constructed a unique risk profile of AI-introduced vulnerabilities:
\begin{enumerate}

\item \textbf{Severity Anthropomorphism}: The severity distribution of vulnerabilities introduced by AI is not significantly different from that of humans, indicating that while learning human coding patterns, it also reproduces the severity characteristics of vulnerabilities introduced by humans.

\item \textbf{Attack Vector Networking}: Vulnerabilities introduced by AI are highly concentrated in network attack vectors, revealing that its generated code is more likely to form security flaws that can be exploited remotely.

\item \textbf{Root Cause Patterning}: Vulnerabilities introduced by AI may stem from the reproduction of insecure coding patterns in the training data, rather than complex logical errors. This explains the phenomenon of vulnerability concentration in high-frequency scenarios such as network programming.

\end{enumerate}

\section{Evolutionary Trends of AIGCode and Its Security Implications}

\begin{figure}[htbp]
    
    \begin{subfigure}{\linewidth} 
        \centering
        \includegraphics[width=0.95\linewidth]{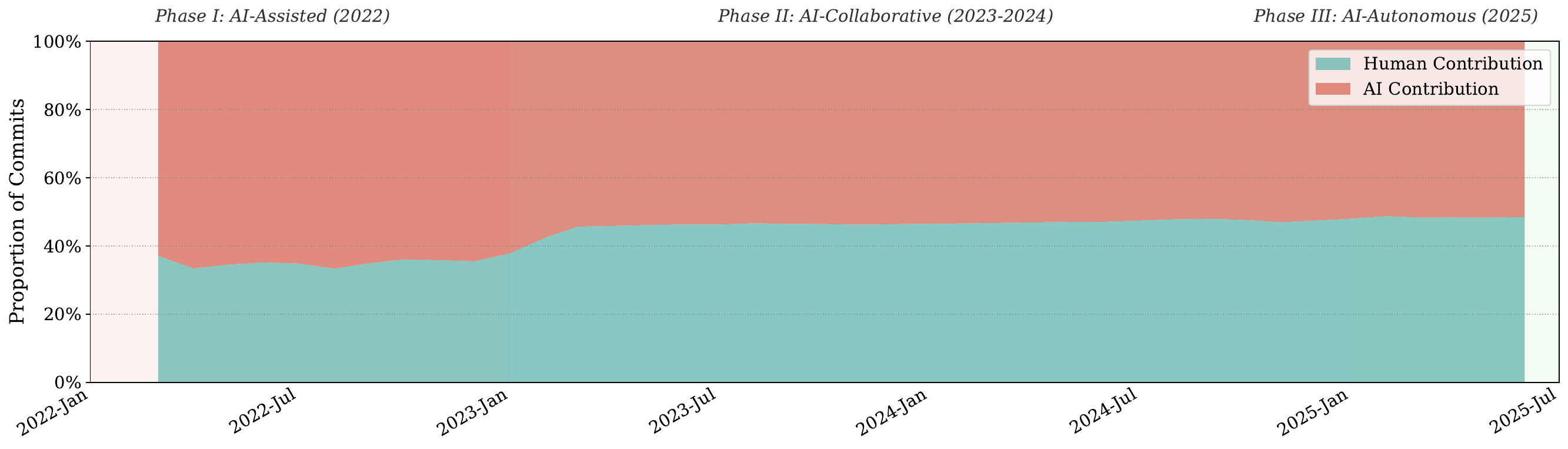}
        \caption{Temporal Trend of AI vs. Human Code Contribution}
        \label{fig:7-1-adoption}
    \end{subfigure}
    
   \begin{subfigure}{\linewidth} 
        \centering
        \includegraphics[width=0.95\linewidth]{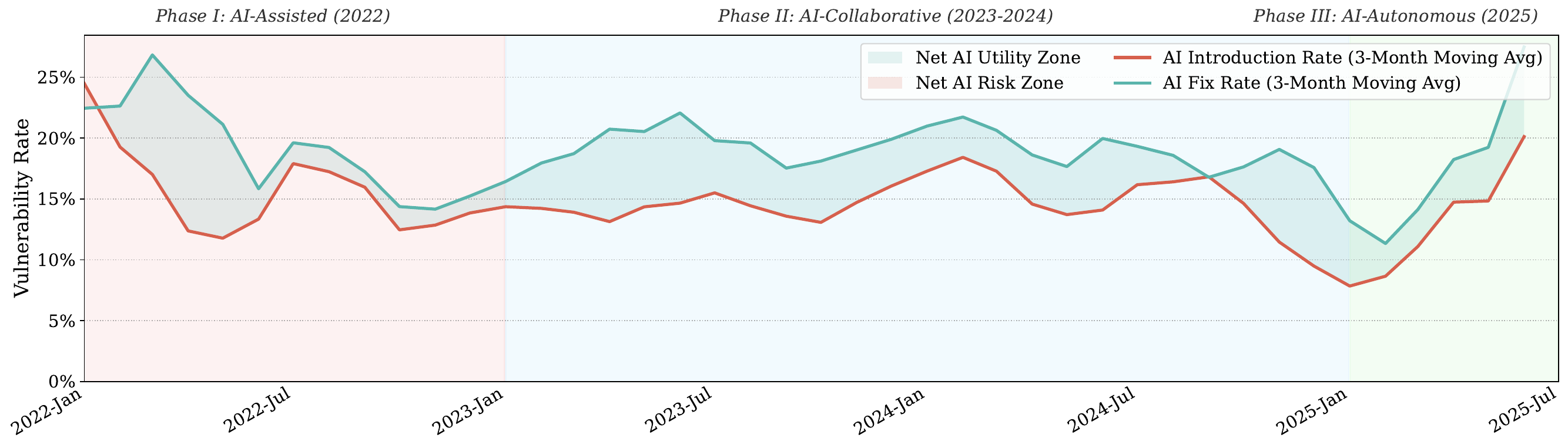}
        \caption{Temporal Dynamics of AI Introduction and Fix Rates}
        \label{fig:7-2-vul rates}
    \end{subfigure}
    
    \begin{subfigure}{\linewidth} 
        \centering
        \includegraphics[width=0.95\linewidth]{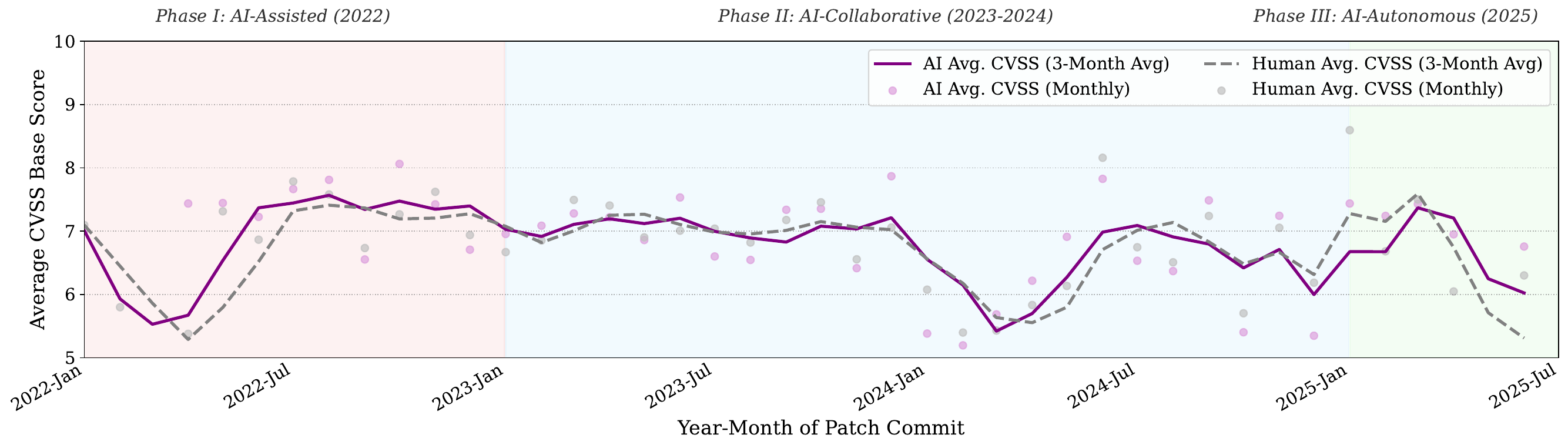}
        \caption{Temporal Comparison of Vulnerability Severity }
        \label{fig:7-3-severity}
    \end{subfigure}
    \centering
    \caption{Temporal Trend Analysis of AI Empowerment: Contribution, Security Role, and Severity}
    \label{fig:7-aigcode_multi_analysis}
\end{figure}

\paragraph{The Early Adoption Phase: Exploring the Utility of AI-Assisted Programming}

In the AI-assisted programming phase marked by the launch of GitHub Copilot, AI quickly established itself as a core tool in the early adoption community. As shown in Fig.\ref{fig:7-aigcode_multi_analysis}(\subref{fig:7-1-adoption}), AI's contribution remained high at 60\%–68\% throughout 2022, indicating its immediate and deep penetration as a code generation tool. This phenomenon is mainly attributed to the high efficiency and accuracy AI demonstrated when handling tasks with abundant training data, such as function completion, code translation, and boilerplate generation.

However, the high adoption rate was accompanied by the early emergence of security risks. Fig.\ref{fig:7-aigcode_multi_analysis}(\subref{fig:7-2-vul rates}) shows that both the introduction and remediation rates of AI vulnerabilities peaked at approximately 25\% at the beginning of the year before declining significantly, both falling back to around 12\% by mid-year. This synchronized decline reflects that after extensive initial exploration, developers began to focus their AI applications on more cautious scenarios with more clearly defined uses.

It is worth emphasizing that AI consistently maintained a net security utility advantage during this phase: its repair rate was consistently 3–5 percentage points higher than its introduction rate, proving that the reliability of AI as a patterned vulnerability repair tool has been validated in practice. However, Fig.\ref{fig:7-aigcode_multi_analysis}(\subref{fig:7-3-severity}) reveals a key risk signal: the average CVSS score of vulnerabilities introduced by AI rebounded rapidly after a brief decline and stabilized in the high-risk range of 7.3–7.5, comparable to the severity level of vulnerabilities introduced by humans. This indicates that even in assisted mode, vulnerabilities introduced by AI are not limited to low-risk defects but possess persistent high-risk characteristics, suggesting that it may systematically reproduce coding anti-patterns with serious consequences contained in the training data.

In summary, the first phase showed a rapid shift from broad exploration to in-depth focus, establishing the net security benefits of AI while also clarifying the potential high severity risks of vulnerabilities it introduces.

\paragraph{The Collaborative Phase: Balancing Risks and Benefits in AI-Human Collaboration}

The AI collaborative programming phase from 2023 to 2024 marks a period of dynamic equilibrium in which AI's role in the vulnerability lifecycle has entered a phase of high participation. As shown in Fig.\ref{fig:7-aigcode_multi_analysis}(\subref{fig:7-1-adoption}), after a gradual decline at the beginning of 2023, the proportion of AI contributions stabilized at a high level of around 55\%, indicating that AI has reached the upper limit of participation under the current collaborative paradigm.

Against this backdrop, security roles have become subtly differentiated. Fig.\ref{fig:7-aigcode_multi_analysis}(\subref{fig:7-2-vul rates}) shows that for most of 2023, the AI fix rate and the introduction rate maintained similar trends, with the fix rate consistently leading; however, in mid-2024, the fix rate fluctuated independently, showing a trend of first rising and then falling, indicating that the enhanced functionality of collaboration tools was used to perform centralized proactive security maintenance.

The key turning point at the end of 2024 further revealed a collective reflection within the industry: both the introduction and fix rates declined significantly, with the introduction rate dropping more dramatically, allowing the fix rate to maintain a relative advantage even at its low point. This trend indicates that, after long-term collaboration, developers strengthened their review and static analysis of AI outputs, effectively curbing the introduction of new vulnerabilities.

Meanwhile, Fig.\ref{fig:7-aigcode_multi_analysis}(\subref{fig:7-3-severity}) reveals an ecosystem-level shift in risk characteristics: in the first half of 2024, the CVSS scores for AI- and human-introduced vulnerabilities exhibited a synchronized ``V-shaped'' fluctuation, strongly suggesting that the entire development ecosystem was influenced by both external events and internal practices. Despite these fluctuations, the high severity of AI-introduced vulnerabilities persisted throughout the phase.

Overall, the second phase is a transitional period for AI from widespread trial to cautious application, reflecting the dynamic balance between tool capabilities and security risks in practice.

\paragraph{The Future Paradigm: Challenges and Prospects of Autonomous AI Programming}

In the autonomous programming paradigm phase from 2025 to the present, the security impact of AI has exhibited significant asymmetric characteristics. As shown in Fig.\ref{fig:7-aigcode_multi_analysis}(\subref{fig:7-1-adoption}), the contribution of AI has stabilized at a saturation level of 65\%, indicating that its share of the total workload has reached the upper limit of the current technological paradigm.

In terms of security, a divergence in capabilities was observed: Fig.\ref{fig:7-aigcode_multi_analysis}(\subref{fig:7-2-vul rates}) shows that both the vulnerability introduction rate and the remediation rate rebounded sharply after hitting a low point at the end of 2024, with the growth rate of the remediation rate significantly exceeding that of the introduction rate for the first time. This trend demonstrates that the new generation of AI agents has made breakthroughs in vulnerability remediation tasks. Their efficiency in autonomous analysis, localization, and generating remediation solutions far exceeds the risk of introducing new errors, making them a powerful provider of net security utility.

Meanwhile, Fig.\ref{fig:7-aigcode_multi_analysis}(\subref{fig:7-3-severity}) reveals a structural shift in risk characteristics: the CVSS scores for vulnerabilities introduced by AI and humans spiked simultaneously in the early stages of the phase before significantly declining. This synchronized decline is more likely due to the large-scale deployment of AI agents to handle high-frequency, automatically identifiable, but low-severity defects, leading to a dilution of the overall average rather than a reduction in the inherent risk.

The characteristics of the third stage can be summarized as an asymmetric explosion of efficiency and a strategic shift in security focus, marking a fundamental evolution of AI's role from a high-level assistant to an autonomous security tool.

\section{Discussion}
Our measurements reveal that AIGCode is not uniformly distributed throughout a codebase; rather, it clusters in highly structured, repetitive regions—such as documentation, tests, and glue code—while only selectively penetrating core logic. This distribution is not merely a curiosity; it serves as a design signal: contemporary programming abstractions are unevenly “AI-friendly.” In domains characterized by regularity, strong typing, and template-like structure, models exhibit competence and developers demonstrate a willingness to delegate. Conversely, in areas where semantics are brittle, underspecified, or encoded in ad-hoc configuration languages, AI adoption declines and error rates rise significantly.

One interpretation of these findings may appear pessimistic—suggesting that AI is not yet ready for the more challenging aspects of software development. A more constructive reading, however, points toward an opportunity: language and API design can explicitly treat AI as a first-class consumer. By intentionally designing interfaces, type disciplines, and configuration formats that are easier for models to reason about, we can foster correct-by-construction behavior not only for humans, but also for AI.

The security implications further sharpen this perspective. The model does not introduce entirely new classes of bugs; rather, it systematically instantiates a narrow set of insecure idioms—such as specific output-encoding lapses or cryptographic misuses. While concerning in practice, this tendency offers a valuable “error signature” for the programming languages and software engineering research communities. It suggests a research agenda that inverts traditional bug-finding: instead of asking “what bugs exist in this program?”, we might ask “what bugs does this model systematically produce?” This shift enables the design of type systems, effect systems, linters, or synthesis constraints that render such error patterns unrepresentable or syntactically unattractive. In this sense, the failure modes of AI can inform the design requirements for future languages and tools.

Our findings on human–AI collaboration reveal another critical tension. Current workflows implicitly assume that human reviewers serve as the final defense: AI drafts, humans review. Yet the scale and velocity of AIGCode render this assumption increasingly fragile. Review capacity does not automatically scale with generation throughput; indeed, larger AI-authored patches tend to receive more superficial scrutiny. This mismatch suggests that the unit of trust in AI-augmented workflows should shift from the individual commit to the entire pipeline. Rather than relying solely on overburdened human reviewers, we should explore combinations of static analysis, differential testing, specification mining, or even ``AI-vs-AI'' adversarial checking to ensure trustworthiness from code suggestion to integration. The central research challenge thus becomes not only how to detect AI code, but where in the development pipeline we can most efficiently embed guarantees—so that AI’s speed becomes an asset, not a liability.

Finally, our study calls into question the default trust model applied to code repositories. Most existing tools and processes are provenance-agnostic: once code is merged into the main branch, it is treated as homogeneous. Our data, however, argue for a more nuanced view: provenance matters. If code originating from AI carries measurably different risk profiles, then ``who—or what—wrote this?'' becomes semantically meaningful. This opens several promising directions for PL and systems research: treating AIGCode as a taint source in static analysis, embedding provenance labels in intermediate representations to prioritize verification efforts, or even elevating the “trust level of origin” to a type- or module-level property. Rather than engaging in binary debates about whether to ``allow'' AI, we can develop language and system mechanisms that encode, refine, and enforce trust boundaries between human and machine authorship within the toolchain itself.

\section{Limitations and Future Work}

\textbf{Our data collection focuses on code commits from 2022 to 2025.} On the one hand, this period corresponds to the most active phase of AIGCode adoption to date; on the other hand, it reflects a practical trade-off between coverage and data processing costs. However, various code completion tools and intelligent programming assistants were already in use before this window. Whether AIGCode from different generations of tools exhibits systematically different styles, distributions, or security risk profiles remains an open question. Taking a longitudinal perspective on this evolution may reveal how AIGCode adoption pathways and risk patterns change over time, and we see this as a promising direction for future work.

\textbf{Several of our core findings methodologically rely on the accuracy of AIGCode detection.} At present, there is no widely accepted, high-precision off-the-shelf tool or model that can be directly applied to large-scale in-the-wild code. Although we design and implement a composite detection scheme that performs well for our setting, it still leaves room for improvement--for example, in more finely identifying AI-originated code that has been heavily edited by humans. Since the primary focus of this paper is on the ecosystem- and security-level impact of AIGCode, rather than on pushing detection performance to its absolute limit, we deliberately do not perform extensive modeling, tuning, or error analysis of the detector itself. Future work can build on our framework to systematically improve detection accuracy and to quantify how detection errors propagate into downstream empirical conclusions, thereby strengthening the robustness of the overall findings.

\section{Conclusion}

Prior to this work, the understanding of AIGCode in real-world development was largely anecdotal or confined to isolated experiments, resulting in a lack of systematic, large-scale empirical insights. To address this gap, we integrate multiple data sources—such as repository commit histories and CVE-linked patches—and deploy a novel AI code detection pipeline. Our findings confirm that AIGCode has become a substantial and distinctive component of modern software development, introducing both productivity gains and novel security challenges.

In this study, we investigate how AIGCode is transforming the software ecosystem and examine the security implications that accompany this shift. Through a large-scale empirical analysis—encompassing top GitHub repositories and thousands of vulnerability-related code changes—we present the first data-driven portrait of AIGCode as it occurs “in the wild”.

\bibliographystyle{ACM-Reference-Format}
\bibliography{software}

\end{document}